\documentstyle{article}
\begin{document}

\pagestyle{empty}
\rightline{UG-64/91}
\rightline{December 1991}
\vspace{2truecm}
\centerline{\bf  $N=2\ W$-supergravity}
\vspace{2truecm}
\centerline{\bf E.~Bergshoeff \footnote
{Bitnet address: bergshoeff@hgrrug5}
and M.~de Roo \footnote
{Bitnet address: deroo@hgrrug5}
}
\vspace{.5truecm}
\centerline{Institute for Theoretical Physics}
\centerline{Nijenborgh 4, 9747 AG Groningen}
\centerline{The Netherlands}
\vspace{3.5truecm}
\centerline{ABSTRACT}
\vspace{.5truecm}
We quantise the classical gauge theory of $N=2\ w_\infty$-supergravity
and show how the underlying
$N=2$ super-$w_\infty$ algebra gets deformed
into an $N=2$ super-$W_\infty$ algebra.
Both algebras contain the $N=2$ super-Virasoro algebra as
a subalgebra. We discuss how
one can extract from these results information about quantum
$N=2\ W_N$-supergravity theories
containing a finite number of higher-spin symmetries
with superspin $s\le N$. As an
example we discuss the case of quantum $N=2\ W_3$-supergravity.
\vfill\eject
\pagestyle{plain}

\noindent{\bf 1. Introduction}

\vspace{.5cm}

In recent years there has been considerable interest in extensions
of the Virasoro algebra containing higher-spin generators. The first
example of such a higher-spin extension, the $W_3$ algebra
\cite{Za}, contains besides the usual Virasoro
spin-2 generator an additional generator of spin-3. Subsequently,
a more general set of so-called $W_N$ algebras, containing higher-spin
generators of spin $2\le s\le N$, were introduced \cite{Fa}.
Further properties of the $W_N$ algebras have been discussed in
\cite{Ba,Bi1}.

One may further generalise the $W_N$ algebras in different ways.
First of all, it is possible to consider $W$-algebras with an infinite number
of higher-spin generators. The first example of such an algebra
is the $w_\infty$ algebra \cite{Bak}. Other examples are the
so-called $W_\infty$ and $W_{1+\infty}$ algebras \cite{Po1,Po2}.

Secondly, one may consider supersymmetric extensions of the
$W$-algebras, both with a finite as well as with an infinite
number of higher-spin generators.
The supersymmetric extension of $w_\infty$ was given in
\cite{Se,Po3}, and of $W_\infty \oplus W_{1+\infty}$ in \cite{Be3}.
The latter algebra can be defined for arbitrary values
of the central charge. It turns out that it is not so easy to
construct a similar super-$W$ algebra with a finite number of generators
(see, e.g.~\cite{In}-\cite{It}). Most examples
of super-$W_N$ algebras\footnote{
We consider here only quantum algebras.
An algebra is called classical with respect to a given field realisation
if the algebra can be realised as a Poisson bracket algebra between
currents which depend on the fields. The algebra is called quantum if,
in order to realise the algebra, one needs to make more than single
contractions between the currents (the single contractions correspond
to the Poisson brackets).}
 given sofar exist only for specific values of
the central charge. In fact, as far as we know only in two cases the
explicit OPE expansions defining a super-$W_N$ algebra have
been given in the literature. These are the $N=2$ super-$W_2$
algebra \cite{Fi,Ko} and the $N=2$ super-$W_3$ algebra
\cite{Ro1}\footnote{The $N=2$ super-$W_3$ algebra seems to be the first
member of a whole family of quantum $N=2$ super-$W_N$ algebras which
can be defined for arbitrary values of the central charge
\cite{Ne,It}.}.
Furthermore, in \cite{Ah} the existence of an $N=1$ supersymmetric
extension of $W_3$ has been argued.

It has by now become clear that the bosonic higher-spin $W$-symmetries occur
in a number of, quite unexpected, places.
To give a few examples, the $W_N$-symmetries play a role in the
context of conformal field theories with $c\ge 1$, exceptional
modular invariants \cite{Ba}, nonlinear differential equations,
Toda theories \cite{Bi1} and matrix models of 2D-gravity \cite{Fu1}.
Similarly, $W_\infty$-symmetries were found in recent studies of
the first Hamiltonian structure of the KP hierarchy \cite{Ya},
matrix models of 2D-gravity \cite{Go,Fu2}, discrete states \cite{Kl},
two-dimensional black holes \cite{El} and string field theories \cite{Av}.

In most of the above examples the presence of the
$W$-symmetries was discovered a posteriori.
One could also consider the $W$-symmetries as
fundamental symmetries and treat them on
the same footing as the Virasoro symmetries. The aim here is
to extend the ordinary string to a so-called ``W-string''.
Recently, several steps in this programme have been undertaken,
including
the gauging of the $W$-symmetries \cite{Hu1}-\cite{Be6},
an investigation of the
anomaly-structure in $W$-gravity
\cite{Hu1},\cite{Ma}-\cite{Po4}, and a study of the
spectrum of $W$-strings \cite{Da,Po7}.

It seems natural to investigate the role of supersymmetry
in the above examples. Supersymmetric $W$-symmetries
were found in, for instance, the super-KP hierarchy \cite{Fig,Yu,Das}.
Again, one
could consider super-$W$ symmetries as fundamental
symmetries underlying a $W$-superstring theory. It is well-known
from ordinary string theory that the additional supersymmetry brings in
attractive features. For instance, it removes the tachyon which is
present in the bosonic string spectrum. It is to be expected that
similar things will happen in the case of $W$-superstrings.

With the above motivation in mind we will investigate in this paper the
structure of $W$-supergravity theories.
We will define our starting point, which is the classical gauge theory of
$N=2\ w_\infty$-supergravity in section 3.
The underlying algebra of this gauge theory is discussed in section 2.
We will use a representation
in which the
matter fields are represented by two scalar superfields,
corresponding to a two-dimensional target space. Ultimately,
our goal is to use a multi-scalar representation corresponding to
a higher-dimensional target space. Along the lines of the advances
which have been made recently in the bosonic case,
we will show in section 4 that the theory can be consistently quantised,
thereby removing all so-called matter-dependent anomalies.
In this process the underlying classical algebra gets deformed
into a quantum algebra, as in the  bosonic case \cite{Be2}. We will exhibit
the structure of the quantum algebra in section 5 and in section 6
discuss the remaining so-called universal anomalies.
Both sections 2 and 5, which deal with the classical and quantum algebra,
respectively, can be read independently of the rest of the paper.

The main part of this paper deals with the case of
$W_\infty$-symmetries. However, as is well known from the bosonic case
\cite{Be1}, one can sometimes truncate a theory with $W_\infty$-symmetries
to a theory with $W_N$-symmetries. We will discuss this point in section 7
and show in which sense our results give information on the structure
of $W_N$-supergravity.
In particular, we will discuss the case of $N=2\ W_3$-supergravity.
Finally, in section 8 we give our conclusions and in the Appendix
we give some representative examples of OPE expansions.

We indicate Planck's constant $\hbar$ explicitly when we want to emphasize
the distinction between classical and quantum aspects.

\vspace{.5cm}

\noindent{\bf 2. The $N=2$ super-$w_\infty$ algebra}

\vspace{.5cm}

The $N=2$ super-$w_\infty$ algebra \cite{Po3}
is a higher-spin extension of the $N=2$ super-Virasoro algebra
with generators $w^{(s)}\ (s=1,3/2,2,\dots)$.
The algebra can be defined
by giving the (singular part of the)
OPE expansions of the generators.
The OPE expansion of two generators $w^{(s)}, w^{(t)}$ where both $s$ and $t$
are integer is given by (we set $\hbar=1$ in this section)

\begin{equation}
w^{(s)}(1)w^{(t)}(2) \sim -2{\theta_{12}w^{(s+t-1/2)}\over z_{12}}
\end{equation}
In all other cases the OPE expansion is given by

\begin{eqnarray}
w^{(s)}(1)w^{(t)}(2) &\sim& (-)^{|2s+1|_2}
\{(s+t-{3\over 2}){\theta_{12}w^{(s+t-3/2)}\over z_{12}^2}\\
&&-{1\over 2} {D_2w^{(s+t-3/2)}\over z_{12}}
+ (s-{1\over 2}){\theta_{12}\partial_2w^{(s+t-3/2)}\over z_{12}}\}\nonumber
\end{eqnarray}
where $|s|_2$ is equal to zero for $s$ even and $1$ for $s$ odd.
Furthermore, we have defined
$z_{12}= z_1-z_2 +\theta_1\theta_2$.
The superspace coordinates are $(Z,\bar Z)=(z,\theta,\bar z,\bar\theta)$.
The superspace differential operators $D,\bar D$ are defined by

\begin{equation}
D={\partial\over \partial\theta}-\theta\partial
\hskip 1.5cm
\bar D = {\partial\over \partial\bar\theta} - \bar\theta\bar\partial
\end{equation}
where $\partial = \partial_z, \bar\partial = \partial_{\bar z}$
(corresponding to a Euclidean-signature on the world-sheet),
$\partial_\theta,\partial_{\bar\theta}$
are left-derivatives and $D^2=-\partial, \bar D^2 = -\bar\partial$.
Note that $D_1z_{12}=D_2z_{12}=-\theta_{12}$
and $D_1\theta_{12}=-D_2\theta_{12}=1$.
We will often use the short-hand notation $w^{(s)}(1)$ to indicate
$w^{(s)}(Z_1,\bar Z_1)$, etc. From eqs.~(1) and (2) one can recover the
commutation relations of the generators of the algebra by multiplying the
OPE's by the parameters of the corresponding transformations and
integrating over the superspace coordinates.

It turns out that it is possible to extend the $N=2$ super-$w_\infty$
algebra with an additional $s=1/2$ generator $w^{(1/2)}$ with\footnote{
In cases where we would like to stress the presence of the $s=1/2$
generator we will call the extended algebra,
in analogy with the terminology $w_\infty$ versus
$w_{1+\infty}$, the $N=2$ super-$w_{1/2+\infty}$ algebra.}
$w^{(1/2)}(1)w^{(1/2)}(2) \sim 0$.
The OPE expansion of $w^{(1/2)}$
with $w^{(s)}$ (s integer) is given by

\begin{equation}
w^{(1/2)}(1)w^{(s)}(2) \sim
{w^{(s-1/2)}\over z_{12}} + \{(s-1){\theta_{12}w^{(s-1)}\over z_{12}^2}
- {1\over 2}{D_2w^{(s-1)}\over z_{12}}\}
\end{equation}
For half-integer $s\ (s\ge 3/2)$, the OPE expansion is given by

\begin{eqnarray}
w^{(1/2)}(1)w^{(s)}(2) &\sim&
\{(s-1){\theta_{12}w^{(s-1)}\over z_{12}^2}
-{1\over 2}{D_2w^{(s-1)}\over z_{12}}\}\ \ + \\
&&\{{1\over 2}(s-{3\over 2}){w^{(s-3/2)}\over z_{12}^2}
-{1\over 4}{\theta_{12}D_2w^{(s-3/2)}\over z_{12}^2}
-{1\over 4}{\partial_2w^{(s-3/2)}\over z_{12}}\}\nonumber
\end{eqnarray}

The $N=2$ super-$w_\infty$ algebra contains an $N=2$ super-Virasoro
subalgebra which is generated by $\{w^{(1)},w^{(3/2)}\}$:

\begin{eqnarray}
w^{(1)}(1)w^{(1)}(2) &\sim&
-2 {\theta_{12}w^{(3/2)}\over z_{12}}\nonumber\\
w^{(3/2)}(1)w^{(1)}(2) &\sim&
\{{\theta_{12}w^{(1)}\over z_{12}^2} - {1\over 2} {D_2w^{(1)}\over z_{12}}
+ {\theta_{12}\partial_2w^{(1)}\over z_{12}}\}\\
w^{(3/2)}(1)w^{(3/2)}(2) &\sim&\{{3\over 2}{\theta_{12}w^{(3/2)}\over z_{12}^2}
-{1\over 2}{D_2w^{(3/2)}\over z_{12}} + {\theta_{12}\partial_2w^{(3/2)}
\over z_{12}}\}\nonumber
\end{eqnarray}
The superfields $w^{(1/2)}$ and $\{w^{(s)},w^{(s+1/2)}\}$
with $s$ integer form $N=2$ multiplets with
respect to the $osp(2,2)$ subalgebra of the $N=2$ super-Virasoro subalgebra.
Here $w^{(1/2)}$ constitutes a so-called $N=2$ scalar multiplet.
The $osp(2,2)$ subalgebra is defined by the $s=1,3/2$ transformations
where the parameters $k_{(1)}, k_{(3/2)}$
which multiply the currents $w^{(1)},w^{(3/2)}$
satisfy the conditions

\begin{equation}
D^3k_{(1)}=D^5k_{(3/2)}=0
\label{eq:osp}
\end{equation}

It is possible to perform different truncations of the $N=2$
super-$w_\infty$ algebra. We first consider the ones that maintain the $N=2$
supersymmetry. It turns out that, for a given positive
integer $M\ge 1$, it is consistent to
retain only the $N=2$ multiplets $\{w^{(s)},w^{(s+1/2)}\}$\ (s integer)
with $s=1+kM, k=0,1,2,\dots.$ We denote this algebra by $N=2$ super-$w_{
\infty/M}$. A similar set of truncations has been discussed in the
bosonic case in the second reference of \cite{Hu1}.
For $M=1$ we recover the original
$N=2$ super-$w_\infty$ algebra. Only for $M=1$, it is possible to
extend the algebra by an $s=1/2$ generator
to an $N=2$ super-$w_{1/2+\infty}$ algebra as indicated above.

We next consider truncations giving algebras with $N=1$ supersymmetry
\cite{Se,Po3}. One possibility
is to retain only the generators $w^{(s)}$ with half-integer $s$.
One can then further truncate the algebra by
keeping only the generators $s=3/2+kM$ with
$k=0,1,2,\dots$ and $M\ge 1$ a given integer. Another possibility is
to keep only the generators $w^{(s)}$ with $s$ even or $s+1/2$ even.
We will denote the latter algebra by $N=1$ super-$w_\infty$.

It is also possible to truncate the $N=2$ super-$w_\infty$ algebra
to a finite set of generators by keeping only the
$N=2$ multiplets $\{w^{(s)},w^{(s+1/2)}\}$ with
$s \le M$ for a given integer $M$, giving an $N=2$ super-$w_M$ algebra.
The OPE's of the $N=2$ super-$w_M$ algebra are given by
those of the $N=2$
super-$w_\infty$ algebra with the restriction that

\begin{equation}
w^{(s)}(1)w^{(t)}(2) \sim 0
\end{equation}
for $s+t-1/2 > M$ if $s,t$ integer and $s+t-3/2 > M$ in all
other cases.
This generalizes a similar truncation that takes place in the bosonic case
\cite{Lu3,Li2}.
For clarity we give the field content of some of the
truncated algebras in the table below.

\vspace{.7cm}
\begin{center}
\begin{tabular}{||l|l||}
\hline
superalgebra & $N=1$\ field\ content\\
\hline
$N=2$\ super-$w_{1/2+\infty}$&1/2,1,3/2,2, $\dots$\\
$N=2$\ super-$w_\infty$&1,3/2,2,5/2, $\dots$\\
$N=2$\ super-$w_{\infty/2}$&1,3/2,3,7/2, $\dots$\\
$N=1$\ super-$w_\infty$&3/2,2,7/2,4, $\dots$\\
$N=2$\ super-$w_M$&1/2,1,3/2,2, $\dots , M,M+1/2$\\
\hline
\end{tabular}
\end{center}

\vspace{.4cm}

{\it Table 1.} \hskip .3cm$N=1$ superfield content of some classical super-$w$
algebras. Each spin-$s$ superfield contains two components with
spin $(s,s+1/2)$.

\vspace{.5cm}

Finally, we note that the bosonic subalgebra of $N=2$ super-$w_{1/2+\infty}\
(N=2$ super-$w_\infty)$ is given by the direct sum $w_{1+\infty}\oplus
w_{1+\infty}\ (w_\infty\oplus w_{1+\infty})$.

\vspace{.5cm}

\noindent{\bf 3. Classical $N=2\ w_\infty$-supergravity}

\vspace{.5cm}

The classical theory of chiral $N=2\ w_\infty$-supergravity\footnote{
For results on the gauging of $W_3$-supergravity theories,
see \cite{Hu1,Bas,Mi}. For the gauging of a super-$W_\infty$ algebra, see
\cite{Be5}. It is interesting in its own right to compare the quantum
theory of the $N=2\ w_\infty$-supergravity theory we consider in this
paper with the quantum theory of the $N=2\ W_\infty$-supergravity theory
of \cite{Be5}. It is not obvious to us what the exact relationship
between the two quantum theories is.
}
that
will form our starting point is described by the action $S=1/\pi
\int d^2Z L$, where $L$ is given by \cite{Be8}\footnote{
Note that we have given here the kinetic term in an off-diagonal
basis. After diagonalisation one ends up with the kinetic terms
for two scalar fields with a relative minus sign. One might therefore
consider these two scalars as the coordinates of a superstring
moving in a $d=2$ target space with Lorentzian signature.}

\begin{equation}
L = D\phi\bar D \bar\phi +
\sum_{s=1/2}^\infty A_{(s)} w^{(s)}
\label{eq:L0}
\end{equation}
The matter is described by two real scalar superfields $\phi,\bar\phi$.
The currents $w^{(s)}\ (s=1/2,1,3/2,2,\dots)$ depend on
the matter fields $\phi,\bar\phi$.
Taking single contractions between these currents
(or, equivalently, taking Poisson brackets) one finds the OPE's corresponding
to the $N=2$ super-$w_\infty$ algebra given in the
previous section. Explicitly, the currents are given by

\begin{eqnarray}
\label{eq:clcurrents}
w^{(s)} &=& (\partial\phi)^{s-1}D\phi D\bar\phi
\hskip 5.1cm (s\ {\rm integer)}\\
w^{(s)} &=&  (\partial\phi)^{s-1/2}D\bar\phi
+ {1\over 2} D\{D\phi (\partial\phi)^{s-3/2}D\bar\phi\}
\hskip 1cm (s\ {\rm half-integer)}\nonumber
\end{eqnarray}
We have also introduced gauge fields $A_{(s)}$. We note that
$A_{(s)},w^{(s)}$ are commuting (anticommuting) for integer
(half-integer) $s$.
The two-point function of $\phi,\bar\phi$ is given by

\begin{equation}
<\phi(Z_1,\bar Z_1)\bar\phi (Z_2,\bar Z_2)>\ = - \hbar\ln z_{12}\bar z_{12}
\end{equation}

The action is invariant under $N=2\ w_\infty$-transformations.
Under a spin-$s$ transformation
with $(Z,\bar Z)$-dependent parameter $k_{(s)}$
the $\phi$ matter field transforms as follows:\footnote{
We use a notation where the one-dimensional (anti-commuting) integration
measure is indicated by $dZ$. The two-dimensional (commuting) integration
measure is denoted by $d^2Z \equiv dZd\bar Z$.
}

\begin{equation}
\delta(k_{(s)})\phi(2) = \hbar^{-1}\sum_{s\ge 1/2}
\oint {dZ_1\over 2\pi i}
k_{(s)}(1)w^{(s)}(1)\phi(2)
\label{eq:transf}
\end{equation}
and similarly for $\bar\phi$. Note that $k_{(s)}$ is commuting (anticommuting)
for half-integer (integer) $s$. Using the explicit form of the
currents $w^{(s)}$ and the useful formula

\begin{equation}
\phi(2) = {1\over 2\pi i} \oint dZ_1 {\theta_{12}\over z_{12}} \phi(1)
\end{equation}
we find

\begin{eqnarray}
\delta\phi &=&
\sum_{s=1/2,3/2,\dots}^\infty \{
k_{(s)}(\partial\phi)^{s-1/2} - {1\over 2} Dk_{(s)}D\phi
(\partial\phi)^{s-3/2}\} + \sum_{s=1,2,\dots}^\infty
k_{(s)}D\phi(\partial\phi)^{s-1}\nonumber\\
\delta\bar\phi &=&
\sum_{s=1/2,3/2,\dots}^\infty \{
-(s-{1\over 2})D\bigl (k_{(s)}(\partial\phi)^{s-3/2}D\bar\phi \bigr )\ \ +
\\
&&{1\over 2} Dk_{(s)}(\partial\phi)^{s-3/2}D\bar\phi
+{1\over 2} (s-{3\over 2})D \bigl (
Dk_{(s)}D\phi(\partial\phi)^{s-5/2}D\bar\phi \bigr )\}\ \ +\nonumber\\
&&\sum_{s=1,2,\dots}^\infty \{
-k_{(s)}(\partial\phi)^{s-1}D\bar\phi
-(s-1)D\bigl (
k_{(s)}D\phi(\partial\phi)^{s-2}D\bar\phi)\}\nonumber
\end{eqnarray}
We note that the variation of the kinetic term in the action cancels
against the inhomogeneous variation $\bar Dk_{(s)}$
of the gauge fields, i.e.

\begin{equation}
\int d^2Z\ \biggl (
\delta\{D\phi\bar D\bar\phi\} + \sum_{s\ge 1/2} \bar D k_{(s)}w^{(s)}
\biggr ) =0
\end{equation}
The remaining variation $\hat\delta A_{(s)}$ of the gauge fields,
defined by $\delta A_{(s)}\equiv\bar D k_{(s)}
+\hat\delta A_{(s)}$,
can be determined by the requirement that

\begin{equation}
\int d^2Z\ \sum_{s\ge 1/2} \{\hat\delta A_{(s)}w^{(s)} +
A_{(s)}\delta w^{(s)}\} =0
\end{equation}
where $\delta w^{(s)}$ may be calculated by applying the
general formula (\ref{eq:transf}) and the OPE expansions of the
$N=2$ super-$w_\infty$ algebra.
We thus obtain the following expression for the
transformation rules of the gauge fields under the spin $s\ge 1$
transformations:

\begin{equation}
\delta A_{(s)} = \bar D k_{(s)} -2 \sum_{t=1,2,\dots}^{s-1/2}
A_{(t)}k_{(s-t+1/2)}
\end{equation}
for half-integer $s$ and integer $t$ and

\begin{eqnarray}
\delta A_{(s)} = \bar D k_{(s)} \ \ +
&&\sum_{t={1\over 2},1,\dots}\{
-(t-{1\over 2}) A_{(t)}\partial k_{(s-t+3/2)}\ \ + \\
&&{1\over 2}(-)^{|2t|_2}DA_{(t)}Dk_{(s-t+3/2)}
+ (s-t+1)\partial A_{(t)}k_{(s-t+3/2)}\}\nonumber
\end{eqnarray}
in all other cases.
It is understood that $k_{(s)}\equiv 0$ if
$s\le 1/2$.
Under the spin $s=1/2$ transformations the gauge fields transform
as follows:

\begin{eqnarray}
\delta A_{(s)} &=& -(s+{1\over 2})A_{(s+1)}\partial k_{1/2}
+{1\over 2}DA_{(s+1)}Dk_{1/2}\\
&&-{1\over 2}(s+1)A_{(s+3/2)}\partial Dk_{1/2}
+{1\over 4}DA_{(s+3/2)}\partial k_{1/2}
-{1\over 4}\partial A_{(s+3/2)}Dk_{1/2}\nonumber
\end{eqnarray}
for integer $s$. For half-integer $s$ we have:

\begin{equation}
\delta A_{(s)} = -A_{(s+1/2)}Dk_{1/2}
-(s+{1\over 2})A_{(s+1)}\partial k_{1/2}
-{1\over 2}DA_{(s+1)}Dk_{1/2}
\end{equation}
On both the matter fields $\phi,\bar\phi$
as well as on the gauge fields $A_{(s)}$ the commutator
algebra of the $k_{(s)}$ transformations closes and corresponds to the
classical $N=2$ super-$w_\infty$ algebra.

We should comment on the gauging of the lowest-spin $s=1/2$ transformation.
At first sight one might be surprised that it is possible to gauge
this transformation. Indeed, in the
one-scalar realisation of the
bosonic $w_\infty$-gravity theory the lowest-spin $s=1$
transformation is not gauged \cite{Be1}.
The reason for this difference is that we are working here with
a two-scalar realisation where the lowest-spin  $s=1/2$ current is
given by $w^{(1/2)}=D\bar\phi$ and, since $<\bar\phi(1)\bar\phi(2)>\ =0$,
a single contraction between two $s=1/2$ currents does not give a central term.
Therefore, one can treat the $s=1/2$ transformation on the same footing
as the higher-spin transformations.
On the other hand, in the usual one-scalar formulation of the
bosonic $w_\infty$-gravity theory the lowest-spin $s=1$ current
is given by $w^{(1)}=\partial\phi$ and, since now $<\phi(1)\phi(2)>\ \ne 0$,
a single contraction between two $s=1$ currents yields a central term,
giving a ``classical anomaly''\footnote{
The gauging of algebras with central charges has been discussed in \cite{Hu6}.
}.

It is instructive to take the bosonic limit of the
$N=2\ w_\infty$-supergravity theory. In this limit
we are left with two real scalars $\phi,\bar\phi$ with the
two-point function given by

\begin{equation}
<\partial\phi(1)\partial\bar\phi(2)>\ = {\hbar\over (z_1-z_2)^2}
\end{equation}
The Lagrangian for these scalars reads

\begin{equation}
L = \partial\phi\bar\partial\bar\phi + \sum_{s=1}^\infty A_{(s)}w^{(s)}
\end{equation}
with the currents $w^{(s)}\ (s=1,2,\dots)$ given by

\begin{equation}
w^{(s)} = (\partial\phi)^{s-1}\partial\bar\phi
\end{equation}
The corresponding Poisson bracket
algebra is the bosonic $w_{1+\infty}$-algebra.
In terms of (the singular part of) the operator product (OPE)
expansion of the currents this algebra is given by

\begin{equation}
\hbar^{-1}w^{(s)}(1)w^{(t)}(2) \sim (s+t-2){w^{(s+t-2)}\over (z_1-z_2)^2}
+ (s-1){\partial_2w^{(s+t-2)}\over z_1-z_2}
\end{equation}
Note that we have ended up with a two-scalar realisation of
$w_{1+\infty}$-gravity. The $s=1$
generator can be truncated consistently, leading to a two-scalar
realisation of $w_\infty$-gravity. In \cite{Be1} a one-scalar realisation
of $w_\infty$-gravity was obtained.

It is interesting to compare the two-scalar realisation of the
$w_\infty$ and $w_{1+\infty}$ algebras we just found with the
two-scalar realisation found recently in \cite{Sh}. It turns out
that there is a whole one-parameter family of two-scalar realisations
of $w_\infty$ with currents given by

\begin{equation}
w^{(s)}=(\partial\phi)^{s-1}\partial\bar\phi
+\alpha {s-2\over s}(\partial\phi)^s
\end{equation}
Indeed, one may verify that for any choice of the parameter $\alpha$
the Poisson-bracket algebra of the above currents is equal to
$w_\infty$. The two-scalar realisation of \cite{Sh}
corresponds to the choice $\alpha=1$. We find that only for $\alpha=0$
a single contraction between two $s=1$ currents does not yield
a central term. Therefore, only for $\alpha=0$
does the inclusion of the $s=1$ current lead to a $w_{1+\infty}$ algebra
without central extension.

The above
realisations of the classical $w_\infty$ algebra cannot be extended to
a realisation at the quantum level (i.e. taking multiple contractions
between the currents) for arbitrary $\alpha$.
{}From \cite{Sh} it is clear that,
without introducing any further fields, this is possible
for $\alpha=1$.
In our case, with $\alpha=0$ and the $s=1$ generator included,
it is also possible but we
have to modify the currents with terms bilinear in fermions.
These are of course exactly the fermions which occur in the
$N=2\ w_\infty$-supergravity theory\footnote{In the supersymmetric case
one is furthermore forced to consider the direct sum $w_{1+\infty}\oplus
w_{1+\infty}$.}.

\vspace{.5cm}

\noindent{\bf 4. Quantisation}

\vspace{.5cm}

We now proceed to quantise the chiral $N=2\ w_\infty$-supergravity theory.
In this section we will closely follow a similar analysis for
the bosonic $w_\infty$-gravity theory \cite{Be2} and the
bosonic $w_3$-gravity theory \cite{Hu5,Sc2}.
Like in the bosonic case we should distinguish between matter-dependent
anomalies and universal anomalies. The first are generated by
supergraphs with external matter fields and are typical for
nonlinearly realised symmetries. The latter correspond to
supergraphs with external gauge fields only. In this section we
will show how the matter-dependent anomalies can be
eliminated from the theory by suitable finite renormalisations of
the supercurrents and the transformation rules.
The universal anomalies will be discussed in section 6.

\vspace{5.5cm}

{\it Fig.~1} \hskip .3cm  Two supergraphs giving rise to matter-dependent
anomalies.

\vspace{1cm}

As an example of a supergraph  that can generate matter-dependent anomalies
we consider the
sample diagrams given in Fig.~1.
These are the only two diagrams that have an external $A_{(1)}$ and $A_{(2)}$
gauge field and one additional external matter field.
The two diagrams can be calculated by evaluating the double contractions in the
operator product expansions of $\int d^2Z_1A_{(1)}(Z_1)w^{(1)}(Z_1)$ times
$\int d^2Z_2A_{(2)}(Z_2)w^{(2)}(Z_2)$.
The resulting contribution to the effective action is

\begin{eqnarray}
\Gamma_{12\phi} &=&  {\hbar\over \pi^2}
\int d^2Z_1d^2Z_2\ \{A_1(Z_1)A_2(Z_2){1\over z_{12}^2}\partial_2\phi
+A_1(Z_1)A_2(Z_2){\theta_{12}\over z_{12}^3}D_2\phi\}\nonumber\\
&=& {\hbar\over \pi}
\int d^2Z_1d^2Z_2\ A_1(Z_1)A_2(Z_2)\{-D_1{\partial_1\over \bar D_1}
\Delta(Z_1-Z_2)\partial_2\phi
+{1\over 2}{\partial_1^2\over \bar D_1} \Delta(Z_1-Z_2) D_2\phi\}
\nonumber\\
&=& -{\hbar\over \pi}
\int d^2Z\ \{({1\over \bar D}\partial D A_1)A_2\partial\phi
+{1\over 2}({\partial^2\over \bar D} A_1)A_2 D\phi\}
\end{eqnarray}
The delta function $\Delta(Z_1-Z_2)$ is defined by

\begin{equation}
\Delta(Z_1-Z_2) = \delta(z_1-z_2)(\theta_1-\theta_2)(\bar\theta_1 -
\bar\theta_2)
\end{equation}
We have furthermore defined a regularisation where

\begin{equation}
{\theta_{12}\over z_{12}} = \pi {1\over \bar D_1}\Delta(Z_1-Z_2)
\end{equation}
The inverse operator $1/\bar D$ is defined by the relations

\begin{equation}
{1\over \bar D} \bar D = \bar D {1\over \bar D} = 1
\end{equation}
By taking repeated derivatives one can derive the general identities

\begin{eqnarray}
{\theta_{12}\over z_{12}^n} &=& \pi {(-)^{n-1}\over (n-1)!}{\partial_1^{n-1}
\over \bar D_1} \Delta(Z_1-Z_2)\\
{1\over z_{12}^n} &=& \pi {(-)^{n-1}\over (n-1)!} D_1 {\partial_1^{n-1}\over
\bar D_1} \Delta(Z_1-Z_2)\nonumber
\end{eqnarray}
Under the leading order inhomogeneous terms in the gauge transformations
$\delta A_{(1)} = \bar D k_{(1)} + \dots, \delta A_{(2)} = \bar D k_{(2)}
+ \dots$ the anomalous variation of $\Gamma_{12\phi}$ is

\begin{eqnarray}
\label{eq:anom}
\delta \Gamma_{12\phi} =
{\hbar\over \pi} \int d^2Z&& \{
(\partial D k_{(1)})A_{(2)}\partial\phi - {1\over 2}(\partial^2k_{(1)})
A_{(2)}D\phi \nonumber\\
&+& (\partial D A_{(1)})k_{(2)}\partial\phi - {1\over 2}
(\partial^2 A_{(1)})k_{(2)}D\phi \}\\
&+& {\rm equation\ of\ motion\ terms}\nonumber
\end{eqnarray}

It turns out that the
anomalous variation (\ref{eq:anom}) can be cancelled by adding
the finite local counter terms $L_{1/2}+L_1$, given by

\begin{eqnarray}
\label{eq:count}
L_{1/2} &=& \sqrt\hbar\biggl ( A_{(1)}\partial\phi + A_{(2)}(
{1\over 3}\partial D\phi D\bar \phi
+ {1\over 2} (\partial\phi)^2 + {1\over 3}\partial\phi\partial\bar\phi
-{1\over 3}D\phi\partial D\bar\phi)\nonumber\\
&+&A_{(5/2)}({1\over 6}\partial^2\phi D\bar\phi + {1\over 3}
\partial D\phi\partial\bar\phi + {1\over 2}\partial D\phi\partial\phi
-{1\over 3}\partial\phi\partial D\bar\phi - {1\over 6}D\phi\partial^2
\bar\phi) \biggr )\nonumber\\
L_1 &=& \hbar \biggl ( {1\over 6}A_{(2)}
\partial^2\phi + {1\over 12}A_{(5/2)}\partial^2 D\phi \biggr )
\end{eqnarray}
and by simultaneously correcting the transformations of the matter fields
$\phi$ and $\bar\phi$ as well as the gauge field $A_{(1)}$ by
extra terms given by

\begin{eqnarray}
\label{eq:corr}
\delta_{1/2}\phi &=& \sqrt \hbar \{
{1\over 3}k_{(2)}\partial D\phi - {1\over 3} D(k_{(2)}\partial\phi)
+ {1\over 3}\partial (k_{(2)}D\phi)\}\nonumber\\
\delta_{1/2}\bar\phi &=& \sqrt\hbar\{ -Dk_{(1)}
 + {1\over 3}k_{(2)}\partial D\bar\phi
-{1\over 3}D(k_{(2)}\partial\bar\phi)
- D(k_{(2)}\partial\phi) + {1\over 3}\partial
(k_{(2)}D\bar\phi)\}\nonumber\\
\delta_{1/2}A_{(1)} &=& -{2\over 3}\sqrt\hbar \{
A_{(2)}(\partial Dk_{(1)}) + (\partial DA_{(1)})k_{(2)}\}
\nonumber\\
\delta_{1}\bar\phi &=& {\hbar\over 6}\partial D k_{(2)}
\end{eqnarray}
Note that the powers of $\hbar$ are in agreement with the fact
that in two dimensions $\phi$ and $\bar\phi$ have the dimension of
$\sqrt\hbar$. In varying the effective action the terms of order
$\sqrt\hbar$ cancel identically:

\begin{equation}
\delta_0L_{1/2} + \delta_{1/2}L_0 \equiv 0
\end{equation}
where $L_0$ is the $\hbar$-independent part of the Lagrangian given in
eq.~(\ref{eq:L0}).
The terms of order $\hbar$ are such that they cancel
the anomalous variation (\ref{eq:anom}). They arise in the pattern

\begin{equation}
\delta_0L_1+\delta_{1/2}L_{1/2}+\delta_1L_0
\end{equation}

The occurrence of the counterterms (\ref{eq:count}) implies that the original
classical currents $w^{(1)},w^{(2)}$ and $w^{(5/2)}$
have received corrections. A similar correction is found to the $w^{(3/2)}$
current if one considers a Feynman diagram with an external
$A_{(3/2)}$ and $A_{(2)}$ gauge field and one external matter field.
At this point one has found the complete corrections to all currents
$w^{(s)}$ up to and including $s=5/2$.
Dimension counting shows that no higher order in $\hbar$ corrections to these
currents are to be expected. The final expressions for the
quantum currents $W^{(s)}\ (1/2\le s \le 5/2)$ are given by

\begin{eqnarray}
\label{eq:qucu}
W^{(1/2)} &=& D\bar\phi\nonumber\\
W^{(1)} &=& D\phi D\bar\phi + \sqrt\hbar\partial\phi\nonumber\\
W^{(3/2)} &=& {1\over 2}\partial\phi D\bar\phi + {1\over 2} D\phi\partial\bar
\phi + {1\over 2}\sqrt\hbar\partial D\phi\nonumber\\
W^{(2)} &=& \partial\phi D\phi D\bar\phi + \sqrt\hbar
\{{1\over 3}\partial D\phi D\bar\phi + {1\over 2}(\partial\phi)^2
\nonumber\\
&+& {1\over 3}\partial\phi\partial\bar\phi
-{1\over 3}D\phi\partial D\bar\phi\} + {\hbar\over 6}\partial^2\phi\\
W^{(5/2)} &=&
{1\over 2}(\partial\phi)^2D\bar\phi - {1\over 2}D\phi\partial D\phi D\bar\phi
+ {1\over 2}D\phi\partial\phi\partial\bar\phi\nonumber\\
&+&
\sqrt\hbar\{{1\over 6}\partial^2\phi
D\bar\phi+{1\over 3}\partial D\phi\partial\bar\phi
+{1\over 2}\partial D\phi\partial\phi\nonumber\\
&-&{1\over 3}\partial\phi\partial D\bar\phi - {1\over 6}D\phi\partial^2\bar\phi
\} + {\hbar\over 12}\partial^2D\phi\nonumber
\end{eqnarray}
We note that there is an arbitrariness in the above expressions
corresponding to the freedom to make redefinitions of the form
$W^{(s)} \rightarrow W^{(s)}+DW^{(s-1/2)}+\dots$. This arbitrariness
can be removed by requiring that the currents transform covariantly
under the $osp(2,2)$ subalgebra of the $N=2$ super-Virasoro algebra.
This leaves us still with a free undetermined parameter $\lambda$\footnote{
The situation is different in the one-scalar realisation of the
bosonic $w_\infty$-gravity theory. In order to avoid the introduction of the
anomalous $s=1$ transformation one must go to a basis with $\lambda=0$
\cite{Be2}.}.
Different choices of this parameter correspond to choosing a
different basis of the quantum algebra.
For simplicitly, we have given the expressions above only in the
$\lambda=0$ basis. It is straightforward to derive the expressions
for arbitrary value of $\lambda$. Results for arbitrary $\lambda$
will be given in the next section.

The transformation rules for the matter fields $\phi$, including the
corrections (\ref{eq:corr}), follow from the standard expression
(\ref{eq:transf}) where one should use now the quantum currents $W^{(s)}$
instead of the classical currents $w^{(s)}$.
Finally, the modified transformation rule for the $A_{(1)}$ gauge field
follows from the fact that the
OPE expansion of the quantum currents $W^{(s)}$ differs from that of
the classical currents. For instance,

\begin{equation}
w^{(1)}(1)w^{(2)}(2) \sim -2{\theta_{12}w^{5/2}\over z_{12}}
\end{equation}
but

\begin{equation}
W^{(1)}(1)W^{(2)}(2) \sim -2{\theta_{12}W^{(5/2)}\over z_{12}}
+{2\over 3}\sqrt\hbar {W^{(1)}\over z_{12}^2}
\end{equation}
The modified transformation rule of the $A_{(1)}$ gauge field
can now be determined from the requirement that \cite{Po4}

\begin{equation}
\int d^2Z\ \sum_{s\ge 1/2}\{\hat\delta A_{(s)}W^{(s)}
+A_{(s)}\hat\delta W^{(s)}\} =0
\label{eq:rule1}
\end{equation}
where $\hat\delta A_{(s)}=\delta A_{(s)}-\bar D k_{(s)}$ and\footnote{
It is understood here that in the OPE expansion $W^{(s)}(1)W^{(t)}(2)$
${\underline {\rm no}}$ central charge terms are included. These central
charge terms will play a role in section 6 where we will discuss the
universal anomalies.}

\begin{equation}
\hat\delta W^{(t)}(2)
= \hbar^{-1}\sum_{s\ge 1/2}\oint {dZ_1\over 2\pi i}
k_{(s)}(1)W^{(s)}(1)W^{(t)}(2)
\label{eq:rule2}
\end{equation}

One can now in principle proceed, by looking at higher-order diagrams
with higher-spin external gauge fields, to determine the appropriate
modifications to the higher-spin currents $w^{(s)}$ with $s\ge 3$
that are needed in order to remove the matter-dependent anomalies.
The quantum-corrected currents are denoted by $W^{(s)}$. The local
part of the effective action is now given by

\begin{equation}
S_{\rm eff}({\rm local}) = {1\over \pi} \int d^2Z\
\{D\phi\bar D \bar\phi + \sum_{s\ge 1/2}A_{(s)}W^{(s)}\}
\label{eq:Slocal}
\end{equation}

At the same time, the transformation rules of the matter and gauge
fields will require higher-spin modifications too. As in the sample
diagrams studied above, the modifications to  the $\phi$ and
$\bar\phi$ variation will be precisely those that follow by
substituting the quantum currents into (\ref{eq:transf}). The
modifications to the gauge field variations follow from (\ref{eq:rule1}).
These constructions can be carried out to arbitrary order in $\hbar$.

We will now argue that, like in the bosonic case, the modifications
to the Lagrangian and transformation rules can all be understood as a
renormalisation of the classical $N=2$ super-$w_\infty$ algebra to the quantum
$N=2$ super-$W_\infty$ algebra.
First of all, since the modifications
to the currents generate the modifications
to the matter fields as in (\ref{eq:transf}) it follows that
all variations

\begin{equation}
\biggl ( \delta_0 + \delta_{1/2} + \delta_1 + \dots \biggr )
D\phi\bar D\bar\phi
\end{equation}
of the kinetic term of the effective action are cancelled by the variation

\begin{equation}
\sum_{s\ge 1/2}
\biggl ( (\delta - \hat \delta)A_{(s)}\biggr )W^{(s)} =
\sum_{s\ge 1/2} \bar D k_{(s)}W^{(s)}
\end{equation}
The remaining variation
of the effective action is calculated as follows.
The variation of the currents $W^{(s)}$ in the
$A_{(s)}W^{(s)}$ terms in (\ref{eq:Slocal}) is given by

\begin{equation}
\delta W^{(t)} = \hat\delta W^{(t)} - ({\rm double\ +\ more\ contractions})
\end{equation}
with $\hat\delta W^{(t)}$ defined in (\ref{eq:rule2}). The second
term at the r.h.s.~indicates that to calculate the variation of
quantum currents one should only consider the contribution of the
the single contractions to the OPE expansion $W^{(s)}(1)W^{(t)}(2)$
in (\ref{eq:rule2}).
Now using (\ref{eq:rule1}) we find that the total variation of
$S_{\rm eff}({\rm local})$ is given by

\begin{equation}
\delta S_{\rm eff}({\rm local}) = -{1\over \pi} \int d^2Z\ \sum_{s\ge 1/2}
A_{(s)}\biggl ( \hat \delta W^{(s)} - ({\rm single\ contractions})\biggr )
\end{equation}
One may verify that this expression is exactly the same (with an opposite sign)
to the contribution that follows from the Feynman diagram calculation.
Each double or more contraction in the calculation
of the quantum algebra corresponds to a particular Feynman diagram.
In other words, in the variation of the effective action, the contribution of
the local part corresponds to the single contractions in the
quantum algebra whereas the contribution from the nonlocal part,
arising from the Feynman diagram calculations, corresponds
to the double + more contractions.
Together they lead to a closed quantum algebra and via
(\ref{eq:rule1}) to an invariant effective action. Therefore, by
construction the cancellation of the matter-dependent
anomalies is equivalent to the construction of a closed quantum
algebra.

Before proceeding with the
cancellation of the universal anomalies in section 6, we will first consider
in the next section some basic properties of the quantum $N=2$ super-$W_\infty$
algebra. We will describe the algebra in terms of a one-parameter
family of bases with parameter $\lambda$ \cite{Be4}.
To compare with the results of this section one should
take the basis corresponding to
$\lambda=0$.

In particular, we will give in the next section a closed
expression for the structure constants
of the $N=2$ super-$W_\infty(\lambda)$ algebra.
Using these structure constants one can give a closed expression
for the quantum corrected transformation rules of the gauge fields
$A_{(s)}$. The explicit form of these transformations
is given in eq.~(\ref{eq:gftransf}).

\vspace{.5cm}

\noindent{\bf 5. The  $N=2$ super-$W_\infty(\lambda)$ Algebra}

\vspace{.5cm}

In this section we describe the structure of the
$N=2$ super-$W_\infty(\lambda)$ algebra. This section
follows the analysis of \cite{Be4} with the notation
adapted to this paper.

The quantum
$N=2$ super-$W_\infty(\lambda)$ algebra can be described as the algebra
formed by arbitrary positive powers of the superspace differential
operator $D$.
The explicit expressions for the differential operators are given by
($s={1\over 2},1,{3\over 2},2,\dots$)
\begin{equation}
L_\lambda^{(s)}(k_{(s)}) = \sum_{i=0}^{2s-1} A^i(s,\lambda)
\bigl (D^{2s-i-1}k_{(s)}\bigr )D^i
\end{equation}
where $\lambda$ is the conformal weight. The summation index $i$ takes
only integer values. The parameter
$k_{(s)}$ is commuting (anticommuting)
for half-integer (integer) $s$.
The algebra formed by the $L^{(s)}$ operators generalize
the super-Virasoro algebra which is generated by

\begin{equation}
L_\lambda^{(3/2)}(k_{(3/2)}) =
-k_{(3/2)}D^2 - {1\over 2}(Dk_{(3/2)})D
-\lambda(D^2k_{(3/2)})
\label{eq:difop}
\end{equation}
The coefficients $A^i(s,\lambda)$ are given by
\cite{Be4}

\begin{eqnarray}
A^i(s,\lambda) &=& (-)^{[s+1/2]+(2s+1)(i+1)+1}
{[s-1/2]!\over [i/2]![s-i/2-1/2]!}\nonumber \\
&&\times {([i/2+1/2]+2\lambda)_{[s]-[i/2+1/2]}\over
([s+i/2+1/2])_{2s-[s+i/2+1/2]}}
\end{eqnarray}
\noindent
where $(a)_n\equiv (a+n-1)!/(a-1)!$ and $[a]$ denotes the integer
part of $a$.
These  coefficients are fixed by the requirement
that the $L^{(s)}$ form nondecomposable representations of the
$osp(1,2)$ subalgebra of the super-Virasoro algebra.
This $osp(1,2)$ subalgebra is generated by the differential operators
$L_\lambda^{(3/2)}$ with the parameters $k_{(3/2)}$ satisfying

\begin{equation}
D^5 k_{(3/2)} = 0
\end{equation}

It is possible to define a field-theoretic representation of the
super-$W_\infty(\lambda)$  algebra in terms of a superconformal $BC$
system.  In this representation we have
two conformal superfields, a commuting field $B$ and an anticommuting
field $C$, with conformal weights $\lambda$ and ${1\over 2} - \lambda$,
respectively. When subjected to their field equations, these fields
decompose according to $B(z,\theta)=\beta(z)+\theta b(z)$ and
$C(z,\theta)=c(z)+\theta\gamma(z)$.
Since $\theta$ has weight $-{1\over 2}$, we find that $b,c,\beta$
and $\gamma$ have conformal weights $\lambda +{1\over 2},
-\lambda + {1\over 2},\lambda$ and $-\lambda +1$. The supersymmetric
action equals \cite{Fr}

\begin{equation}
S={1\over \pi}\int d^2Z B\bar D C = {1\over \pi}\int
d^2 z\{\beta \bar\partial\gamma + b\bar\partial c\}
\end{equation}

\noindent
The generators of the super-$W_\infty(\lambda)$ algebra are
related to the following conserved supercurrents
of (quasi-)conformal spin $s={1\over 2},1,{3\over 2},2,\dots $:

\begin{eqnarray}
\label{eq:currents}
W_\lambda^{(s)} &=&
\sum_{i=0}^{2s-1}(-)^{[s-i/2]+|2s+1|_2|2s-i-1|_2}
A^i(s,\lambda)D^{2s-i-1}\{(D^iB)C\}\nonumber\\
&=&\sum_{i=0}^{2s-1} {\tilde A}^i(s,\lambda)
(D^iB)(D^{2s-i-1}C)
\end{eqnarray}
\noindent
with the coefficients ${\tilde A}^i(s,\lambda)$ given by \cite{Be4}

\begin{eqnarray}
{\tilde A}^i(s,\lambda) &= &
{1+|2s|_2|i+1|_2\over 1+|2s|_2}
{(-)^{[s]+[{i\over 2}]+|2s+1|_2
|i+1|_2 }\over (-[-s])_{[s]-|2s|_2|i|_2}}
 \pmatrix {[s]-1+|2s|_2|i+1|_2\cr [{i\over 2}]\cr} \nonumber \\
&&\times (2\lambda - [s])_{[{i\over 2}]+|2s+1|_2|i|_2}\ \ \ \
(-2\lambda-[s]+1)_{[s]-[{i\over 2}]-|i|_2}
\end{eqnarray}
In eq.~(\ref{eq:currents})
a normal ordering with respect to the modes of $B,C$ is understood.
The supercurrent $W_\lambda^{(s)}$ is commuting (anticommuting)
for integer (half-integer) $s$.
Each supercurrent contains a spin $s$
and a spin $s+{1\over 2}$ component current. The normalization
of the supercurrents is taken such that the current $W_\lambda^{(s)}$
exactly generates the variation corresponding to the differential
operator\footnote{
For simplicity we will set $\hbar=1$ everywhere in this section.}
$L_\lambda^{(s)}$:

\begin{equation}
\delta(k_{(s)})B(2) = L_\lambda^{(s)}(k_{(s)})B(2) =
{1\over 2\pi i}\oint dZ_1 k_{(s)}(1)W_\lambda^{(s)}(1)B(2)
\end{equation}
with the two-point function of $B$ and $C$ given in eq.~(\ref{eq:OPE}).
The coefficients ${\tilde A}^i(s,\lambda)$ can also be determined by
the requirement that the
supercurrents  $W_\lambda^{(s)}$ form $N=1$ superfields with respect to
the $osp(1,2)$ subalgebra of
the super-Virasoro algebra which is generated by
$W_\lambda^{(3/ 2)}$. We note that the ${\tilde A}^i(s,\lambda)$
satisfy the identity

\begin{equation}
{\tilde A}^i(s,\lambda) = (-)^{-[-s]-1+|2s|_2|i|_2}
{\tilde A}^{2s-i-1}(s,{1\over 2} -
\lambda)
\label{eq:identity}
\end{equation}
{}From this identity we see that the value $\lambda =1/4$ is special.
We find that for that value of $\lambda$ the
coefficients ${\tilde A}^i(s,1/4)$ can be written as

\begin{equation}
{\tilde A}^i(s,{1\over 4}) = {(-)^{|2s|_2+[{i\over 2}+{1\over 2}]+1}\over
(1+|2s|_2) 2^{2[s]-1}}\pmatrix{2s-1\cr i}
\end{equation}

It turns out that the currents
$\{W_\lambda^{(1)},W_\lambda^{(3/2)}\}$
form an $N=2$ super-Virasoro algebra \cite{Fr}. All currents fit into $N=2$
supermultiplets with respect to the $osp(2,2)$ subalgebra of this
$N=2$ super-Virasoro algebra.
This $osp(2,2)$ subalgebra is generated by the differential operators
$\{L_\lambda^{(1)}, L_\lambda^{(3/2)}\}$ with the parameters
$k_{(1)}, k_{(3/2)}$ satisfying

\begin{equation}
D^3k_{(1)}=D^5k_{(3/2)}=0
\end{equation}
This generalises eq.~(\ref{eq:osp}) to the quantum algebra.
The resulting $N=2$ combinations are
$\{W_\lambda^{(s)},W_\lambda^{(s+1/2)}\}$ for integer $s$ and
$W_\lambda^{(1/2)}$, where $W_\lambda^{(1/2)}$ constitutes a
so-called $N=2$ scalar multiplet.

The two-point function of the superfields $B(Z)$ and $C(Z)$
is equal to \cite{Fr}

\begin{equation}
<C(Z_1)B(Z_2)>\  = {\theta_{12}\over z_{12}}
\label{eq:OPE}
\end{equation}
The
$N=1$ super-Virasoro algebra generated by $W_\lambda^{(3/2)}$
is defined by the following operator product expansion:

\begin{equation}
W_\lambda^{(3/2)}(1)W_\lambda^{(3/2)}(2) \sim
{3\over 2}{\theta_{12}W_\lambda^{(3/2)}\over z_{12}^2}
-{1\over 2}{D_2W_\lambda^{(3/2)}\over z_{12}}
+{\theta_{12}\partial_2 W_\lambda^{(3/2)}\over z_{12}}
+2{\lambda -{1\over 4} \over z_{12}^3}\ \ + \ \ {\rm regular}
\end{equation}

\noindent
Here we have used the following super-Taylor expansion

\begin{equation}
W(1) = W(2) +\theta_{12}D_2W + z_{12}\partial_2 W
+\theta_{12}z_{12}\partial_2D_2W + {1\over 2}z_{12}^2\partial_2^2W + \dots
\end{equation}
The OPE expansion of two general supercurrents $W_\lambda^{(s)}$ and
$W_\lambda^{(t)}$ is given by the following expression:\footnote{
For a number of representative cases the explicit form of the
OPE expansions is given in Appendix A.}

\begin{equation}
W_\lambda^{(s)}(1)W_\lambda^{(t)}(2) \sim
\sum_{u={1\over 2}}^{s+t-{1\over 2}} f_{st}^u(D_1,D_2;\lambda)
{\theta_{12}W_\lambda^{(s+t-u)}(2)\over z_{12}}
+ c(s,t;\lambda){\theta_{12}^{|2(s+t)|_2}\over (z_{12})^{s+t +
{1\over 2}|2(s+t)|_2}}
\label{eq:strc}
\end{equation}
The structure functions $f_{st}^u(D_1,D_2;\lambda)$ are polynomials
in the supercovariant derivatives of  degree $2u-1$:

\begin{equation}
f_{st}^u(D_1,D_2;\lambda) = f^u_{st}(\lambda)
\sum_{i=0}^{2u-1} M_{st}^u(i) D_1^iD_2^{2u-i-1}
\end{equation}
The functions $M^u_{st}(i)$ are  fixed by the requirement
of $osp(1,2)$ covariance.
They are given by\footnote{
Strictly speaking the statement is that the functions\\
\centerline{
$\tilde M^u_{st}(i) = (-)^{[i/2+1/2]+[u-i/2]+|2s+1|_2|i|_2}M^u_{st}(i)$
}
are fixed by the $osp(1,2)$ covariance. The extra $i$-dependent
sign factor arises from the particular way we have rewritten the
commutator-algebra calculation of \cite{Be4} in terms of the above OPE
expansions.}

\begin{eqnarray}
\label{eq:Mfunctions}
M_{st}^u(i) &=& (-)^{i(2s+2u+1) + [u-i/2]+|2s+1|_2|i|_2}
{[u-1/2]!\over [i/2]![u-i/2-1/2]!}\\
&&\times ([2s-u+1/2])_{[u-i/2-1/2]+|2u+1|_2|2u-i-1|_2}
\  ([2t-u+1/2])_{[i/2]+|2u+1|_2|i|_2}\nonumber
\end{eqnarray}
The structure constants $f^u_{st}(\lambda)$
can be explicitly computed and are given by the following expression
\cite{Be4}:

\begin{equation}
f_{st}^u(\lambda) = F_{st}^u(\lambda) + (-)^{[-u-1/2]+4(s+u+1)(t+u+1)}
F_{st}^u(1/2-\lambda)
\end{equation}
with

\begin{eqnarray}
F_{st}^u(\lambda) &=& (-)^{2s + |2s|_2|2t+1|_2
|2t+1|_2|2u-1|_2+|2s+2t+2u+1|_2}
\nonumber\\
&&\times (-)^{[s+t-u-1/2]}{(2s+2t-2u-1)!\over (2s+2t-[u+1/2]-1)!}\\
&&\times \sum_{i=0}^{2s-1}\sum_{j=0}^{2t-1}\delta(i+j-2s-2t+2u+1)
A^i(s,{1\over 2}-\lambda)A^j(t,\lambda)(-)^{2i(s+t-u+1/2)}
\nonumber
\end{eqnarray}
Finally, the central charge
$c(s,t;\lambda)$ is given by \cite{Be4}

\begin{eqnarray}
c(s,t;\lambda) &=& \sum_{i=0}^{2s-1}\sum_{j=0}^{2t-1}
(-)^{|i|_2|2s-i+j|_2+|j|_2+[j/2+1/2]+[t-j/2]} \\
&&\times \{|2(s+t)|_2+|2(s+t)+1|_2 \bigl (
|2s+1|_2|i+j+1|_2+|2s|_2|i+j|_2 \bigr ) \} \nonumber\\
&&\times \bigl ( [j/2]+[s-i/2-1/2]+|2(s+t)|_2|2s-i+1|_2|j|_2 \bigr )!
\nonumber\\
&&\times \bigl ( [i/2]+[t-j/2-1/2]+|2(s+t)|_2|2t-j+1|_2|i|_2 \bigr )!
\nonumber\\
&&\times {\tilde A}^i(s,\lambda){\tilde A}^j(t,\lambda)
\nonumber
\end{eqnarray}
In particular we find $c(3/2,3/2;\lambda)\equiv -c/6 = 2(\lambda-1/4)$,
or

\begin{equation}
c=-12(\lambda-1/4)
\end{equation}
where c is the usual central charge parameter of the Virasoro algebra.
This Virasoro algebra is defined by the expansion $W_\lambda^{(3/2)}
= 1/2iG + \theta T$ such that $T$ satisfies the standard Virasoro
algebra

\begin{equation}
T(1)T(2) \sim {2T\over (z_1-z_2)^2} + {\partial_2 T\over z_1-z_2}
+ {c/2\over (z_1-z_2)^4}
\end{equation}
Note that for $\lambda =0$ we have $c=3$ as one would expect for a
supersymmetric $BC$ system. Other choices of $\lambda$
give other choices of the central charge but they refer to other
Virasoro subalgebras.
Note that the $BC$ system is a particular $c=3$
representation\footnote{
Actually, as will be explained later in this section, the $BC$ system
provides us with a $c=3$ as well as with a $c=-3$ representation
of the general algebra.}
 of a
more general class of algebras with an arbitrary central charge parameter
$c$ \cite{Be3}, which occurs linearly in all central terms.
Therefore, in
the $\lambda=0$ basis the central charges $c(s,t)$ of the higher spins are
related to $c$ as follows:

\begin{equation}
c(s,t) = - {1\over 6} {c(s,t;0)\over c(3/2,3/2;0)}c
\end{equation}

{}From the above expressions one can derive some general properties of
the structure constants and the central charges.
First of all, we find

\begin{equation}
f_{3/2s}^{1/2}(\lambda) =
f_{3/2s}^{1}(\lambda) =
f_{3/2s}^{2}(\lambda) =
f_{3/2s}^{3/2}(\lambda) = 0
\end{equation}
in agreement with the $osp(1,2)$ covariance of the equations.
The maximum value $s_{{\rm max}}$ of the spin arising at the
r.h.s.~of the OPE between two currents of spin $s$ and $t$
($s,t \ne 1/2)$ is given
by

\begin{eqnarray}
s_{{\rm max}} &=& s+t-{1\over 2} \hskip 2cm (s,t\ {\rm integer})\\
s_{{\rm max}} &=& s+t-{3\over 2} \hskip 1.9cm\ ({\rm all\ other\ cases})
\nonumber
\end{eqnarray}
These maximum values of the spin can be understood from
the fact that within the so-called wedge subalgebra one can use
the addition rules for the spins according to the $osp(1,2)$ algebra.
This wedge subalgebra can be defined by the following restrictions on the
parameters \cite{Be4}:

\begin{equation}
D^{4s-1}k_{(s)} = 0
\end{equation}

Finally, from the general formulae given above one can deduce that
the structure
constants and the central charges satisfy the following identities:

\begin{eqnarray}
\label{eq:sym}
f^u_{st}(\lambda) &=& (-)^{[-u-{1\over 2}] +4(s+u+1)(t+u+1)}
f^u_{st}({1\over 2}-\lambda) \\
c(s,t;\lambda) &=& (-)^{[s+t] + |2s|_2|2t|_2}
c(t,s;\lambda) = - c(t,s;{1\over 2} - \lambda)\nonumber
\end{eqnarray}

On the basis of these relations one can show that the super-$W_\infty
(\lambda)$ and super-$W_\infty(1/2-\lambda)$
algebras (${\underline {\rm without}}$ the central
terms) are equivalent to each other. To be precise, the form
of the OPE expansions (\ref{eq:strc}) does not change if one
replaces $\lambda$ everywhere by $1/2-\lambda$ and furthermore
redefines the currents with a factor $(-)^{[s]+1}$, i.e.,

\begin{equation}
 W_\lambda^{(s)} \rightarrow  (-)^{[s]+1} W_{1/2-\lambda}
^{(s)}
\label{eq:map}
\end{equation}
This equivalence is only true
at the level of single contractions or, equivalently,
Poisson brackets. It ceases to be true if one includes the central
terms in (\ref{eq:strc}), which correspond to
double contractions. In fact, one finds that under the map
(\ref{eq:map}) all the central terms change sign. This change
of sign can be understood as follows. Under the map $\lambda
\rightarrow 1/2-\lambda$ one effectively interchanges the role
of the bosonic $\beta\gamma$ system and the fermionic $bc$ system in the
action.
Both before as well as after this interchange one can
realize the same super-$W_\infty$
algebra (i.e.~with identical structure constants). However,
since bosons have been interchanged with fermions, the
contribution to the central charge changes sign.

As an example,
consider the $BC$ system at $\lambda=0$. On this system
one can realize a $c=3$ representation
of the $N=2$ super-$W_\infty(0)$ algebra. The statement now is that on
a $BC$ system with $\lambda=1/2$ one can realize
a $c=-3$ representation of exactly the same $N=2$
super-$W_\infty(0)$ algebra.

Based on the relations given above one can discuss different truncations
of super-$W_\infty(\lambda)$.
We will briefly discuss here two truncations that are possible at
$\lambda =0$ (or, equivalently, $\lambda = 1/2$) and $\lambda = 1/4$,
respectively.

For $\lambda =0$ one can reduce the
super-$W_\infty(\lambda)$ algebra to an algebra with $N=1$ supersymmetry.
For that value of $\lambda$ one can do two special things. First of all, it
turns out that for $\lambda =0 $ the $s=1/2$ generator can be truncated
away from the algebra.
The reason for this is that for $\lambda = 0$ the $B$ superfield
only occurs as $DB$ in the expressions for the supercurrents, except
in the $s=1/2$ supercurrent. Therefore, in the OPE of two currents
$W_0^{(s)}$ and $W_0^{(t)}$ with $s,t \ne 1/2$ the superfield $B$
will only occur as $DB$. For general values of $\lambda$ this property
is reflected in the fact that in the r.h.s.~of the OPE
the $s=1/2$
supercurrent $W_\lambda^{(1/2)}$ always is multiplied by the second-order
Casimir $C_2$ of $osp(1,2)$ which is $C_2 = \lambda(\lambda-1/2)$.
Consequently, for $\lambda=0$ the $s=1/2$ supercurrent
can be consistently truncated away from the algebra, giving the
$N=2$ super-$W_\infty$ algebra of \cite{Be3}.

Secondly, for $\lambda=0$ the superfields $C$ and $DB$ both have conformal
weight $s=1/2$. One can therefore perform a further truncation of the algebra
by identifying $C$ with $DB$:

\begin{equation}
C \equiv DB
\end{equation}
Note that this identification can only be implemented after having
discarded the $s=1/2$ generator since only then the $B$ superfield
will always occur as $DB$. The effect of the above truncation is that
all supercurrents $W_0^{(s)}$ with $s$ or $s+1/2$ odd vanish identically.
One is then left with the supercurrents $W_0^{(s)}$ with $s$ or
$s+1/2$ even only. They generate
an algebra with $N=1$ supersymmetry which we will
denote with $N=1$ super-$W_\infty$.
The expressions for the currents are given by

\begin{equation}
W_0^{(s)} = \sum_{i=1}^{[s]}{\tilde A}^i(s,0)(D^iB)(D^{2s-i}B)
- {1\over 2} |2s+1|_2{\tilde A}^s(s,0)(\partial^{s/2}B)^2
\end{equation}
We have chosen here the normalization of the currents such that
the nonzero structure constants are exactly the same as the ones of the
$N=2$ super-$W_\infty(0)$ algebra.
We use here the following two-point function for the scalar superfield $B$:

\begin{equation}
<DB(Z_1)B(Z_2)>\  = {\theta_{12}\over z_{12}}
\end{equation}
The expressions $\tilde c(s,t)$ for
the central charges, however, are twice as small, i.e.

\begin{equation}
\tilde c(s,t) = {1\over 2} c(s,t;0)
\end{equation}
In particular, we find that $\tilde c(3/2,3/2)=3/2$ as one would expect
for a single real scalar superfield $B$.

It is interesting to note that the $\lambda=0$ truncation, described
above, is the beginning of a whole series of truncations that take
place for $\lambda=0,-1/2,-1,-3/2,\dots$\footnote{
A similar set of truncations in the bosonic case has been discussed
in the second reference of \cite{Ya}, and
from a different point of view in \cite{Bakas}.}. For instance, for $\lambda=
-1/2$ one can first truncate away the $s=1/2$ generator as well as the
$\{W_{-1/2}^{(1)},W_{-1/2}^{(3/2)}\}$ multiplet. This leads to
an $N=2$ algebra that starts with the $\{W_{-1/2}^{(2)},W_{-1/2}^{(5/2)}\}
$ multiplet. One can then truncate the $N=2$ supersymmetry to
an $N=1$ supersymmetry by making the identification

\begin{equation}
C=\partial DB
\end{equation}
Note that for $\lambda = -1/2$ indeed the conformal weights of $C$ and
$\partial DB$ coincide. The general pattern is then as follows.
For $\lambda=-M/2\ (M=0,1,2,\dots)$ one can truncate away the $N=2$
multiplets $W_{-M/2}^{(1/2)},\{W_{-M/2}^{(1)},W_{-M/2}^{(3/2)}\},
\dots ,\{W_{-M/2}^{(M)},W_{-M/2}^{(M+1/2)}\}$. The truncation to
$N=1$ supersymmetry is then achieved by making the identification

\begin{equation}
C=\partial^M DB
\end{equation}
Note that such identifications lead to higher-derivative actions for the
$B$ superfield:

\begin{equation}
S={1\over \pi}\int d^2Z (\bar D B) \partial^M DB
\end{equation}
Since all these truncations
(except for $M =0)$ lead to algebras that
do not contain a Virasoro subalgebra
we will not consider them further in this paper.

In contrast to the $\lambda=0$ truncation, the $\lambda=1/4$ truncation
preserves the $N=2$ supersymmetry of the super-$W_\infty(\lambda)$ algebra.
On the basis of the symmetry properties of the structure constants
given in eq.~(\ref{eq:sym}) one deduces that the structure
constants $f^u_{st}(1/4)$ vanish identically, whenever
$[-u-1/2]+4(s+u+1)(t+u+1)$ is odd. This enables one to show that
for $\lambda = 1/4$ one can perform a consistent truncation of the
super-$W_\infty(1/4)$ algebra such that one retains
the $\{W_{1/4}^{(s)}, W_{1/4}^{(s+1/2)}\}$ $N=2$
supermultiplets with $s$ odd only. This truncated algebra is related to the
symplecton higher-spin superalgebra of \cite{Bie,Fra}.
We note that the classical version of the truncated $N=2$
super-$W_\infty(1/4)$
algebra is the $N=2$ super-$w_{\infty/2}$ algebra which we introduced in
section 3.

We have shown that the
quantum $N=2$ super-$W_\infty(\lambda)$ algebra
can be truncated for $\lambda=0$ and $\lambda=1/4$. In section 3 we have
discussed similar truncations of the classical $N=2$ super-$w_\infty$ algebra.
We should stress that every truncation of the quantum algebra corresponds
to a truncation of the corresponding classical algebra but that the
reverse is not true: the classical algebra allows truncations that have no
quantum analogue. In the table below we have given the
classical limits of the $\lambda=0$ and $\lambda=1/4$ truncated
quantum algebras discussed above.

\vspace{.7cm}

\begin{center}
\begin{tabular}{||l|l||}
\hline
classical\ algebra&quantum\ algebra\\
\hline
$N=2$ super-$w_{1/2+\infty}$&$N=2$ super-$W_\infty(\lambda)$\\
$N=2$ super-$w_\infty$&$N=2$ super-$W_\infty$\\
$N=2$ super-$w_{\infty/2}$&$N=2$ super-$W_\infty(1/4)$\\
$N=1$ super-$w_\infty$&$N=1$ super-$W_\infty$\\
\hline
\end{tabular}
\end{center}

\vspace{.4cm}

{\it Table 2.} \hskip .3cm Truncations of some classical $w_\infty$
superalgebras and their quantum extensions.

\vspace{.5cm}

\noindent
With the OPE expansions given in the Appendix one may verify the
consistency of the truncations.

We should note that in the generic case the currents $W_\lambda^{(s)}$
are quasi-primary but not primary
with respect to the $N=1$ super-Virasoro algebra
generated by $W_\lambda^{(3/2)}$.
Only the currents $W_\lambda^{(1)},W_{1/4}^{3/2}$ and $W_0^{(2)}$\ (or
$W_{1/2}^{(2)}$) are $N=1$ primary.
The reason for this
is that we preferred to work with a realisation of the currents
in terms of bilinears of the $B,C$ superfields.
In this realisation the $N=2$ super-$W_\infty(\lambda)$ algebra is a linear
algebra. On the other hand,
in most of the literature on nonlinear $W$-algebras a basis is used where all
generators are primary. In our case, we also could have used a
primary basis but, to represent the currents,
we should allow not only bilinears in $B,C$ but also terms
which are quadrilinear and of higher order in $B,C$.
In the primary basis the super-$W_\infty(\lambda)$ algebra
is nonlinear.

To illustrate the above point, we consider the first two
currents beyond the $N=2$ super-Virasoro algebra, i.e.~$W_\lambda^{(2)}$ and
$W_\lambda^{(5/2)}$. These currents are given in terms
of bilinears of $B,C$ (see Appendix A). The current $W_\lambda^{(2)}$
is only $N=1$ primary for $\lambda=0,1/2$, whereas $W_\lambda^{(5/2)}$
is not primary for any value of $\lambda$.
We will now show how, by allowing also terms of higher-order in
$B$ and $C$, one
can construct currents $W_\lambda^{(2)'}$ and $W_\lambda^{(5/2)'}$,
which are $N=1$ primary for any value of $\lambda$\footnote{A similar
discussion in the case of a bosonized $bc$ system has been given in
\cite{Bou}.}.
Starting from the most general polynomial in $B$ and $C$ we find the
following expressions for $W_\lambda^{(2)'}$ and $W_\lambda^{(5/2)'}$:

\begin{eqnarray}
\label{eq:red}
W_\lambda^{(2)'} &=&
+\alpha (2\lambda-1)(\partial DB)C\nonumber\\
&&+\alpha (2\lambda +1) (\partial B)DC\nonumber\\
&&+ \bigl ( \alpha +2\beta(3\lambda-2)\bigr ) (DB)\partial C\nonumber\\
&&+\beta(6\lambda-1)B\partial DC\nonumber\\
&&+2\bigl ( \alpha\lambda -\beta(3\lambda-2)\bigr )B(DB)CDC\nonumber\\
&&+\bigl ( \alpha\lambda(4\lambda +1) -\beta(12\lambda^2-11\lambda
+1)\bigr )
/ 2(2\lambda-1) B^2 (DC)DC\nonumber\\
&&+\bigl (\alpha\lambda +\beta(3\lambda-1)\bigr ) B^2C\partial C\nonumber
\\
W_\lambda^{(5/2)'} &=&
+(-1+2\lambda)(-1+3\lambda)(-3+4\lambda)(\partial^2 B)C\\
&&+2(-1+3\lambda)(-3+4\lambda)(\partial DB)DC\nonumber\\
&&-2(-5+34\lambda-56\lambda^2+24\lambda^3)(\partial B)\partial C\nonumber\\
&&+(-1+6\lambda)(DB)\partial DC\nonumber\\
&&-\lambda(-1+6\lambda)(-7+12\lambda)B\partial^2 C\nonumber\\
&&+2(-1+2\lambda)(2-23\lambda+24\lambda^2)B(\partial B)CDC\nonumber\\
&&-2(-1+2\lambda)(-4+3\lambda)(-1+4\lambda)B(DB)C\partial C\nonumber\\
&&+2(-1-2\lambda +6\lambda^2)B(DB)(DC)DC\nonumber\\
&&+4\lambda(-1-2\lambda+6\lambda^2)B^2(DC)\partial C\nonumber\\
&&+6\lambda(-1+2\lambda)(-1+6\lambda)B^2C\partial DC
\nonumber
\end{eqnarray}
where $\alpha$ and $\beta$ are two arbitrary parameters. Note that the
$B^2(DC)DC$ term in the expression for $W_\lambda^{(2)'}$
is singular for $\lambda=1/2$. For that value of $\lambda$ the
expression for $W^{(2)'}$ is given by

\begin{eqnarray}
\label{eq:red1/2}
W_{1/2}^{(2)'} &=& +\alpha'\biggl (
(\partial B)DC + (DB)\partial C - B\partial DC \biggr )\nonumber\\
&&+ \beta' B^2 (DC)DC
\end{eqnarray}
with $\alpha'$ and $\beta'$ arbitrary.

The above expressions for $W_\lambda^{(2)'}$ and $W_\lambda^{(5/2)'}$
are not necessarily primary with respect to the $N=2$ super-Virasoro
algebra generated by $\{W_\lambda^{(1)},W_\lambda^{(3/2)}\}$. It turns
out that this is only the case for $\lambda=1/2$
and $\alpha'=2\beta'=-2$. In particular,
it is not true for $\lambda=0$. We expect that the following picture
extends to the higher-spin generators
but we have not proven this. One can define $N=1$ primary currents
$W_\lambda^{(s)'}$ for arbitrary values of $\lambda$. Only for
particular values of $\lambda$ can one define $N=2$ primary
superfields $\{W_\lambda^{(s)}, W_\lambda^{(s+1/2)}\}$\ ($s\ge 2$ integer).
We note that the $N=2$ superfield $W_\lambda^{(1/2)}$ plays a special role.
{}From the OPE expansions given in the Appendix it is clear that,
without redefining the $N=2$ super-Virasoro generators, one
cannot define an $N=1$ or $N=2$ primary current with spin $s=1/2$\footnote{
We thank J.~de Boer for a discussion on this point.}.

The quasi-primary currents $\{
W_\lambda^{(2)}, W_\lambda^{(5/2)}\}$ and the
$N=1$ primary currents $\{W_\lambda^{(2)'}, W_\lambda^{(5/2)'}\}$
are related to each other by means
of a nonlinear redefinition
of the generators of the $N=2$ super-$W_\infty(\lambda)$ algebra.
In general, this redefinition involves the
$s=1/2$ generator. One can show however that for $\lambda=1/2$
and $\alpha'=2\beta'=-2$ the $s=1/2$
generator is absent. The nonlinear redefinitions for this case are given in
eq.~(\ref{eq:qpp}).

In order to make contact with the results  of section 4
where we quantised the $N=2\ w_\infty$-supergravity theory
one should
replace
the $B,C$ superfields by two scalar superfields $\phi,\bar\phi$
by applying the
superbosonization rules \cite{Mar}

\begin{equation}
B=e^\phi \hskip 2truecm C=e^{-\phi}D\bar\phi
\end{equation}
Using these superbosonisation rules one can show that the
quantum currents described in this section in terms of
higher-derivative bilinears in $B$ and $C$ are equivalent
to nonlinear expressions in terms of $D\phi, D\bar\phi$ and
supercovariant derivatives thereof. For instance,
one may verify that for $\lambda=0$
one exactly finds, starting from the
$BC$ currents given in  eq.~(\ref{eq:currents})
the quantum currents given in eq.~(\ref{eq:qucu}). For more details
we refer to \cite{Be8}.

The structure constants of the $N=2$ super-$W_\infty(\lambda)$
algebra can also be used to give a closed expression for the
quantum-corrected transformation rules of the gauge fields $A_{(s)}$.
These transformation rules follow from eqs.~(\ref{eq:rule1}) and
(\ref{eq:rule2}). In applying eq.~(\ref{eq:rule2}) we now use the
structure constants of the full quantum $N=2$ super-$W_\infty(\lambda)$
algebra. The final result is given by

\begin{equation}
\delta A_{(s)} = \bar D k_{(s)}
-\sum_{u,t={1\over 2}}^{s+u-{1\over 2}} A_{(t)}k_{(s-t+u)}
\tilde f_{s-t+u,t}^u (\buildrel \leftarrow \over D_k,
\buildrel \leftarrow \over D_A;\lambda)
\label{eq:gftransf}
\end{equation}
Here it is understood that $k_{(s)}=0$ for $s\le 0$. Furthermore
the derivatives $\buildrel \leftarrow \over D_k\ (\buildrel
\leftarrow \over D_A)$ act on $k_{(s-t+u)}\ (A_{(t)})$ only.
The $\tilde f$ structure constants are given by

\begin{equation}
\tilde f_{st}^u(\buildrel \leftarrow \over D_1,
\buildrel \leftarrow \over D_2;\lambda) =
f_{st}^u(\lambda)\sum_{i=0}^{2u-1}
(-)^{[i/2+1/2]+[u-i/2]+|2u-1|_2+|2s+1|_2}
M_{st}^u(i) \buildrel \leftarrow \over D_1^i
\buildrel \leftarrow \over D_2^{2u-i-1}
\end{equation}
We note that the explicit $i$-dependent sign factors in this equation are
due to the fact that we work with the functions $M_{st}^u(i)$
instead of the $\tilde M_{st}^u(i)$ (see also the footnote before
eq.~(\ref{eq:Mfunctions}).

\vspace{.5cm}

\noindent{\bf 6. Universal Anomalies}

\vspace{.5cm}

Having cancelled the matter-dependent anomalies
in section 4, we discuss in this section
the universal anomalies. These anomalies arise
from diagrams with only external gauge fields.
We will see that the universal anomalies are related to the central terms in
the quantum algebra. Since the central terms are numbers, not containing
any quantum currents, the cancellation of the universal anomalies
requires a mechanism different from that of the matter-dependent anomalies.
We have already seen that the matter-dependent anomalies
can be cancelled by an appropriate renormalization of the gauge field
transformation rules. In this section we will see that the cancellation of the
universal anomalies requires finding a $c=0$ representation of the
quantum algebra in terms of matter and ghosts fields.

Following \cite{Ger,Be2}
we first derive an anomalous Ward identity for the universal anomalies.
Considering only diagrams with external gauge fields, the effective action is,
in terms of operator expectation values,

\begin{equation}
e^{-\Gamma(A_{(s)})} =
\biggl < \exp \bigl ( - {1\over \pi}\int \sum_s A_{(s)}W^{(s)} \bigr )
\biggr >
\end{equation}
Varying this equation with respect to $A_{(s)}(Z_1)$ and differentiating
with respect to $\bar Z_1$, one finds

\begin{equation}
\bar D_1 {\delta\Gamma\over \delta A_{(s)}(Z_1)} =
 {1\over \pi} \biggl < \bar D_1 W^{(s)}(Z_1)\exp \bigl (
-{1\over \pi}\int\sum_t A_{(t)}W^{(t)} \bigr ) \biggr > e^\Gamma
\end{equation}
Using the OPE expansion of the super-$W_\infty(\lambda)$ algebra, we
may calculate

\begin{eqnarray}
&&\bar D_1 W^{(s)}(Z_1)\exp \biggl ( -{1\over \pi}\int
\sum_t A_{(t)}W^{(t)}\biggr )  \\
&&= -{1\over \pi}\int d^2Z_2 A_{(t)}(Z_2)\bar D_1
\biggl ( \sum_{u={1\over 2}}^{s+t-{1\over 2}}
f^u_{st}(D_1,D_2;\lambda){\theta_{12}W^{(s+t-u)}(2)\over z_{12}}
+ c(s,t;\lambda){\theta_{12}^{|2(s+t)|_2}\over
(z_{12})^{s+t+{1\over 2}|2(s+t)|_2}} \biggr )\nonumber\\
&&\times (-)^{|2t|_2|2s+1|_2}\ {\rm exp}\bigl (
-{1\over\pi}\int\sum_t A_{(t)}W^{(t)} \bigr )
\nonumber
\end{eqnarray}
Since $\bar D_1 {\theta_{12}\over z_{12}}=\pi\Delta(Z_1-Z_2)$, we
may perform the $Z_2$ integration.
If we now multiply the whole equation with the factor

\begin{equation}
\sum_{s\ge 1/2} \int d^2 Z_1 (-)^{|2s|_2} k_{(s)}(1)
\end{equation}
we find the following anomalous Ward identity for $N=2\ W_\infty$-supergravity:

\begin{equation}
\delta_k\Gamma = \sum_{s,t \ge {1\over 2}}
{\hat c(s,t;\lambda)\over \pi} \biggl <
\int d^2 Z\ k_{(s)} D^{2(s+t)-1} A_{(t)} \biggr >
\end{equation}
where $\hat c(s,t;\lambda)$ is related to the central charge
$c(s,t;\lambda)$ as follows:

\begin{equation}
\hat c(s,t;\lambda) = {(-)^{
|2s+1|_2|2t|_2}
\over
(s+t+{1\over 2}|2(s+t)|_2-1)!} c(s,t;\lambda)
\end{equation}
In deriving this equation we have used that the
transformation rule of $A_{(s)}$ is given by
(\ref{eq:gftransf}).
Thus we see that the effective action is not invariant under spin-$s$
super-$W_\infty$ transformations, on account of the anomalous terms
on the right-hand side. We note that these terms are exactly the ones
that arise from calculating the central charges in the quantum algebra.
Thus, every central term in the quantum algebra corresponds via
the above expression to a
universal anomaly.

One might hope that
the universal anomalies can be cancelled by integrating over all
(component)
higher-spin gauge fields. In general, this integration gives rise to ghosts
which contribute to the central charge in the Virasoro sector as follows:

\begin{equation}
c_{\rm gh} = \lim_{N\to \infty}\sum_{s={1\over 2}}^N c_{\rm gh}(s)
\end{equation}
with

\begin{equation}
c_{\rm gh}(s)=2(-)^{2s+1}(6s^2-6s+1)
\end{equation}
As discussed in \cite{Ya2,Po6}, one may
define the above (divergent) sum by using an appropriate
zeta function regularisation. Using this regularisation one can calculate
$c_{\rm gh}$ for the different $W_\infty$ algebras.
In particular, in \cite{Po6}
it was found that for the $N=2$ super-$W_\infty$
algebra (without the $s=1/2$ generator\footnote{
We conjecture that, if the $s=1/2$ generator is included,
the ghost contribution to the central charge vanishes.})
in the $\lambda=0$ basis the ghost contribution to the
central charge in the Virasoro sector is given by

\begin{equation}
c_{\rm gh}=3
\end{equation}

The idea is now to cancel this ghost contribution by
an equal (but with an opposite sign) contribution $c_{\rm matter}$
of the matter fields such that

\begin{equation}
c_{\rm total} = c_{\rm gh} + c_{\rm matter} =0
\end{equation}
Following the discussion below eq.~(\ref{eq:sym}) we see that indeed we
can achieve this by taking a $BC$ system in the
$\lambda=1/2$ basis\footnote{
Note that this is consistent with the fact that only for $\lambda=0,1/2$
one can close the quantum algebra without the $s=1/2$ generator.}
such that

\begin{equation}
c_{\rm matter} = -3
\end{equation}
and hence $c_{\rm total}= 3-3 = 0$. Note that using a $BC$ system
in the $\lambda=0$ basis,
which has $c_{\rm matter}=+3$, would not work. We therefore conclude that
the remarkabe anomaly cancellation
which was found in the bosonic $w_\infty$-gravity theory \cite{Be2}
also takes place in the supersymmetric case.

\vspace{.5cm}

\noindent{\bf 7. Truncations}

\vspace{.5cm}

It is known that in the bosonic $w_\infty$-gravity theory there
exists a so-called telescoping procedure which enables one
to truncate the theory to a classical $w_N$-gravity theory
containing only a finite number of higher-spin generators \cite{Be1}.
This procedure requires that some of the
higher-spin currents can be expressed as products of
lower-spin currents, and
makes use of the specific representation
of the $w_\infty$ algebra.
Consider a multi-scalar realisation in
which the currents are given by \cite{Be1}

\begin{equation}
w^{(s)}=  {1\over s}{\rm tr}\ (\partial\phi)^s\hskip 2cm s\ {\rm integer}
\end{equation}
where the trace is in the fundamental representation of $U(N)$ or
$SU(N)$. One may verify that these currents satify the bosonic
$w_\infty$ algebra.
For $s\ge N+1$ one runs out of independent Casimir invariants
and therefore it is possible to write all currents $w^{(s)}$ with $s \ge N+1$
as products of the currents $w^{(s)}$ with $s\le N$. An
extreme case is the one-scalar realisation of $w_\infty$ \cite{Be1} where all
currents can be expressed in terms of the $s=2$ current.

We have seen that in order to supersymmetrise $w_\infty$-gravity
we are forced to work with a two-scalar realisation
of the bosonic $w_\infty$ algebra. Together
with their fermionic partners they form the components of the
two scalar superfields $\phi$ and $\bar\phi$.
In this section we will discuss the
telescoping procedure in this realisation, and discuss the corresponding
mechanism at the quantum level.
The first example of a truncation
beyond ordinary (spin-2) supergravity would give an $N=2\ W_3$-supergravity
theory. We will be mainly dealing with this case
as a specific example.

At first sight it looks that there is no telescoping procedure
for the realisation we are working with. From the expressions
(\ref{eq:clcurrents}) for the classical currents, we deduce that
all currents are linear in $\bar\phi$. Hence there is no way
to write a single current as the product of
lower-spin currents. However, it turns out that nonlinear relations,
which do not contain terms linear in the
currents, are possible.

We will first discuss the situation using an
arbitrary $\lambda$ basis. After that we will restrict ourselves to
the case where the $s=1/2$ current can be consistently truncated
away. As we have seen in section 5, this forces us to
use the $\lambda=0$ or $\lambda=1/2$ basis. We will see that the two different
choices of $\lambda$ lead to inequivalent results.
Our strategy is to first consider identities
between the classical currents $w^{(s)}$ and then to consider their
quantum extension, giving identities between the quantum currents
$W^{(s)}$.

In discussing identities beween the classical
currents $w^{(s)}$ we should distinguish between
those which hold independently of the specific representation one is
using and those which are representation-dependent. As an example of
representation-independent identities
we give here the following set

\begin{equation}
w^{(s)}w^{(t)}-(-)^{|2s|_2|2t|_2}w^{(t)}w^{(s)} = 0
\label{eq:trivial}
\end{equation}
Note that these include the relations $w^{(s)}w^{(s)}=0$ for
half-integer $s$.
It turns out that in our specific representation the above set
of identities can be replaced by the following stronger
conditions:

\begin{eqnarray}
w^{(s)}w^{(t)}&=&0 \hskip 4.8cm s,t\ge 1\nonumber\\
w^{(1/2)}w^{(s)}&=&0\hskip 4.8cm s\ {\rm integer}\\
w^{(1/2)}w^{(s)}&=& {1\over 2} w^{(1/2)}Dw^{(s-1/2)}
\hskip 2cm s\ {\rm half-integer}
\nonumber
\end{eqnarray}
There are more identities, which involve
derivatives of the currents.

To explain the situation in a general $\lambda$ basis
we consider all representation-dependent
identities up to a
total spin $s=2$. We find that the independent relations are given by

\begin{eqnarray}
\label{eq:indepl}
w^{(1/2)}w^{(1)}&=&0\nonumber\\
w^{(1/2)}w^{(3/2)}&=&{1\over 2}w^{(1/2)}Dw^{(1)}\\
w^{(1)}w^{(1)}&=&0\nonumber
\end{eqnarray}
Of course at any spin one can find dependent identities,
either by taking derivatives of lower-spin identities or
by multiplying a lower-spin identity with a current. In addition
to (\ref{eq:indepl}), we find the following dependent identity:

\begin{equation}
Dw^{(1/2)}w^{(1)}-w^{(1/2)}Dw^{(1)}=0
\label{eq:depl}
\end{equation}
The above classical relations cannot be used to express
a higher-spin current in terms of a product of lower-spin currents.
Therefore the classical telescoping procedure as discussed in \cite{Be1}
does not exist in this realisation.

The situation changes drastically in the quantum case. In fact, we will
now show that
a quantum telescoping procedure does exist.
It is to be expected that, when quantising the $N=2\ w_\infty$-supergravity
theory, the above classical relations receive quantum corrections
proportional to Planck's constant $\hbar$.
Of course, one should also replace the classical currents by the quantum
currents. For instance, the representation-independent
identities given in (\ref{eq:trivial}) deform into the
following quantum identities \cite{Ba}\footnote{
We thank K.~Schoutens for a discussion on this point.}

\begin{equation}
(W^{(s)}W^{(t)})-(-)^{|2s|_2|2t|_2}(W^{(t)}W^{(s)}) =
\sum_{1,2,\dots} {(-)^{r+1}\over r!}\partial^r\{W^{(s)}W^{(t)}\}_r
\end{equation}
where $\{W^{(s)}W^{(t)}\}_r$ is defined by the following terms
in the OPE expansion:

\begin{eqnarray}
W^{(s)}(1)W^{(t)}(2) &=& \sum_{r=1,2\dots} {\{W^{(s)}W^{(t)}\}_r(2)\over
z_{12}^r} + (W^{(s)}W^{(t)})(2)\ +\ \dots\nonumber\\
&&+\ \theta_{12}-{\rm dependent\ terms}
\end{eqnarray}
and

\begin{equation}
(W^{(s)}W^{(t)})(2)={1\over 2\pi i}\oint dZ_1{\theta_{12}\over z_{12}}
W^{(s)}(1)W^{(t)}(2)
\end{equation}
denotes the product of the supercurrents $W^{(s)}$ and $W^{(t)}$, normal
ordered with respect to the modes of $W^{(s)}$ and $W^{(t)}$.
We note that the chain rule for derivatives also applies to the
normal ordered product:

\begin{equation}
D(W^{(s)}W^{(t)})=(DW^{(s)}W^{(t)})+(-)^{|2s|_2}(W^{(s)}DW^{(t)})
\end{equation}
All the quantum identities discussed sofar are representation-independent,
and
therefore are valid for $\underline {\rm arbitrary}$ value
of the central charge $c$.

We will now discuss the quantum extension of the representation-dependent
identities\footnote{
To prove these identities, one
may either use a representation in terms of $\phi,\bar\phi$
or $B,C$.}.
We find the following quantum extension of the independent
classical identities (\ref{eq:indepl})\footnote{
For simplicitly, we will omit the subindex $\lambda$ of the quantum
currents everywhere in this section.}:

\begin{eqnarray}
\label{eq:qindepl}
\sqrt\hbar W^{(3/2)} &=&(W^{(1/2)}W^{(1)}) +2\lambda\sqrt\hbar
(W^{(1/2)}DW^{(1/2)})\nonumber\\
&&+{1\over 2}\sqrt\hbar DW^{(1)}-2\lambda\hbar\partial W^{(1/2)}
\nonumber\\
\sqrt\hbar W^{(2)} &=& {1\over 2}(W^{(1)}W^{(1)})
-{1\over 3}(1-4\lambda)\sqrt\hbar (W^{(1)}DW^{(1/2)})\nonumber\\
&&+{1\over 6}(1-4\lambda)\hbar\partial W^{(1)}
+{1\over 3}(1+2\lambda)\hbar (W^{(1/2)}DW^{(1/2)})\\
&&-{2\over 3}\lambda(1+2\lambda)\hbar (W^{(1/2)}\partial W^{(1/2)})\nonumber
\\
&&-{2\over 3}\lambda(1-\lambda)\hbar (DW^{(1/2)}DW^{(1/2)})\nonumber\\
&&+{2\over 3}\lambda(1-\lambda)\hbar^{3/2}\partial DW^{(1/2)}\nonumber\\
(W^{(1/2)}W^{(3/2)})&=&{1\over 2}(W^{(1/2)}DW^{(1)})-2\lambda\sqrt\hbar
(W^{(1/2)}\partial W^{(1/2)})\nonumber
\end{eqnarray}
Of course, one can also extend the dependent identity (\ref{eq:depl}) to the
quantum level where it reads as follows:

\begin{eqnarray}
(W^{(1)}DW^{(1/2)})-(W^{(1/2)}DW^{(1)}) &=&\sqrt\hbar DW^{(3/2)}
-2\lambda\sqrt\hbar (DW^{(1/2)}DW^{(1/2)})
\nonumber\\
&&-2\lambda\sqrt\hbar
(W^{(1/2)}\partial W^{(1/2)})
\\
&&+{1\over 2}\sqrt\hbar\partial W^{(1)} +2\lambda\hbar\partial DW^{(1/2)}
\nonumber
\end{eqnarray}

We see that for $\hbar\ne 0$ the first two identities in (\ref{eq:qindepl})
can be used to solve for $W^{(3/2)}$ and $W^{(2)}$ in terms of
(products of) lower-spin currents.
We therefore conclude that at the
quantum level there exists a telescoping procedure.
It seems very suggestive that this telescoping mechanism extends to all
other higher-spin generators with $s\ge 2$ as well, but we have not proven
this.

The above identities can be used to obtain representations of
nonlinear algebras.
Such a construction was performed for the bosonic $W_N$ algebras in
\cite{Lu1} and for the $N=1$ super-$W_2$ algebra in \cite{Be7}.
The idea is to reinterpret some of the generators of the
$N=2$ super-$W_\infty(\lambda)$ algebra as composite operators
instead of independent generators. This is done
by applying decomposition rules of the type given above to the right-hand side
of the OPE expansions corresponding to the super-$W_\infty(\lambda)$
algebra. A linear algebra with an infinite number of generators
is thus truncated to a nonlinear algebra with a finite number of generators.

{}From now we restrict ourselves to nonlinear algebras not containing
an $s=1/2$ generator.
We have seen in section 5 that the $s=1/2$ generator is a
quasi-primary generator that, without redefining the $N=2$ super-Virasoro
generators, cannot be made primary.
We will
postpone a study of nonlinear algebras involving a spin-1/2 generator
to future work. Given the above
restriction we are forced to work either in the $\lambda=0$
or in the $\lambda=1/2$ basis.
Only for these two choices of $\lambda$ can the $s=1/2$ generator
be truncated away consistently from the $N=2$ super-$W_\infty(\lambda)$
algebra.
In section 5 we have seen that at the
level of Poisson brackets or single contractions the two choices
of $\lambda$ are equivalent.
This equivalence ceases to be true when multiple contractions are
taken into account. Indeed, in section 5 we saw that the $\lambda=0\
(\lambda=1/2)$ basis
provides us with a $c=3\ (c=-3)$ representation of the $N=2$ super-$W_\infty
$ algebra.
We will see that in the case of the nonlinear algebras
multiple contractions already occur in non-central terms in the
OPE's. Consequently, the $\lambda=0$
and $\lambda=1/2$
cases lead to inequivalent nonlinear algebras.
In both cases we will discuss the first example
beyond the $N=2$ super-Virasoro algebra, corresponding to an
$N=2$ super-$W_3$ algebra.

We first consider, both for $\lambda=0$ as well as $
\lambda =1/2$, identities between the
classical currents of the $N=2$ super-$w_\infty$ algebra.
To discuss the case of the $N=2$ super-$W_3$ algebra, it suffices to
consider all representation-dependent relations up to a total spin $s=7/2$.
The independent relations are given by

\begin{eqnarray}
\label{eq:indep}
w^{(1)}w^{(1)} &=&0\hskip 2cm
w^{(1)}w^{(3/2)} =0\nonumber\\
w^{(1)}w^{(2)} &=& 0\hskip 2cm
w^{(1)}w^{(5/2)} =0\\
w^{(3/2)}w^{(2)} &=&0\hskip 2cm
Dw^{(1)}w^{(2)} =0\nonumber
\end{eqnarray}
Below we give the quantum extensions of the above classical identities
for the cases $\lambda=1/2$ and $\lambda=0$ separately.

\vspace{.5cm}

\noindent{\bf A. The $\lambda=1/2$ case}

\vspace{.5cm}

It turns out that for $\lambda=1/2$ there are no representation-dependent
quantum identities
at spin $s=2$ and $s=5/2$ not involving an $s=1/2$ generator. The
quantum extension of the independent identities at $s=3$ and $s=7/2$
are given by

\begin{eqnarray}
\label{eq:qindep1/2}
\sqrt\hbar W^{(3)} &=& {2\over 3}(W^{(1)}W^{(2)})
+{4\over 9}\sqrt\hbar (W^{(1)}DW^{(3/2)})\nonumber\\
&&+{2\over 3}\sqrt\hbar (W^{(3/2)}DW^{(1)}) -{1\over 5}\hbar DW^{(5/2)}
-{2\over 15}\hbar^{3/2}\partial^2W^{(1)}\\
\sqrt\hbar W^{(7/2)} &=&
2(W^{(3/2)}W^{(2)}) +{4\over 3}\sqrt\hbar (W^{(3/2)}DW^{(3/2)})\nonumber\\
&&+{2\over 5}\hbar\partial DW^{(2)}-{1\over 6}\hbar^{3/2}\partial^2 W^{(3/2)}
\nonumber
\end{eqnarray}
and

\begin{eqnarray}
\label{eq:nonl1}
{2\over 3}(W^{(1)}DW^{(2)}) - {4\over 3}(DW^{(1)}W^{(2)}) &=&
{4\over 9}\sqrt\hbar (W^{(1)}\partial W^{(3/2)})
+{2\over 9}\sqrt\hbar (DW^{(1)}DW^{(3/2)})\nonumber\\
&&-{2\over 3}\sqrt\hbar
(W^{(3/2)}\partial W^{(1)})\nonumber\\
&&+\hbar\partial W^{(5/2)} -{1\over 3}\hbar^{3/2}\partial^2
DW^{(1)}\\
-(W^{(1)}W^{(5/2)}) +{8\over 3}(W^{(3/2)}W^{(2)}) &=&
-{1\over 3}\sqrt\hbar
(W^{(1)}\partial DW^{(1)}) -{4\over 9}\sqrt\hbar (W^{(3/2)}DW^{(3/2)})
\nonumber\\
&&+{1\over 3}\sqrt\hbar
(DW^{(1)}\partial W^{(1)})\nonumber\\
&&-{2\over 3}\hbar\partial DW^{(2)}
+{4\over 9}\hbar^{3/2}\partial^2 W^{(3/2)}\nonumber
\end{eqnarray}
Furthermore, there is one dependent identity at spin $s=7/2$:

\begin{eqnarray}
\label{eq:nonl2}
{2\over 3}(W^{(1)}DW^{(2)})+{2\over 3}(DW^{(1)}W^{(2)}) &=&
\sqrt\hbar DW^{(3)} - {10\over 9}\sqrt\hbar (DW^{(1)}DW^{(3/2)})
\nonumber\\
&&+{4\over 9}\sqrt\hbar (W^{(1)}\partial W^{(3/2)})
-{2\over 3}\sqrt\hbar (W^{(3/2)}\partial W^{(1)})\nonumber \\
&&-{1\over 5}\hbar \partial W^{(5/2)}
-{1\over 30}\hbar^{3/2}\partial^2 DW^{(1)}
\end{eqnarray}

\vspace{.5cm}

\noindent{\bf B. The $\lambda=0$ case}

\vspace{.5cm}

We find for $\lambda=0$ the following independent quantum identities:

\begin{eqnarray}
\label{eq:qindep0}
\sqrt\hbar W^{(2)} &=& {1\over 2}(W^{(1)}W^{(1)})
-{1\over 3}\hbar DW^{(3/2)}\nonumber\\
\sqrt\hbar W^{(5/2)} &=& (W^{(1)}W^{(3/2)})-
{1\over 6}\hbar\partial DW^{(1)}\nonumber\\
\sqrt\hbar W^{(3)} &=& {2\over 3}(W^{(1)}W^{(2)})
+{2\over 9}\sqrt\hbar (W^{(1)}DW^{(3/2)})-
{2\over 5}\hbar DW^{(5/2)}\\
\sqrt\hbar W^{(7/2)} &=& 2(W^{(3/2)}W^{(2)})
+\sqrt\hbar (W^{(1)}\partial DW^{(1)})\nonumber\\
&&+{2\over 3}\sqrt\hbar(W^{(3/2)}DW^{(3/2)})
+{1\over 2}\sqrt\hbar (\partial W^{(1)}DW^{(1)})\nonumber\\
&&-{2\over 5}\hbar\partial DW^{(2)}-{2\over 3}\hbar^{3/2}\partial^2
W^{(3/2)}\nonumber
\end{eqnarray}
and

\begin{eqnarray}
\label{eq:qindep}
(W^{(3/2)}W^{(2)})-{1\over 2}(W^{(1)}W^{(5/2)}) &=&
-{5\over 12}\sqrt\hbar (W^{(1)}\partial DW^{(1)})
-{1\over 3}\sqrt\hbar(W^{(3/2)}DW^{(3/2)})\nonumber\\
&&+{1\over 4}\hbar^{3/2}\partial^2W^{(3/2)}\\
(DW^{(1)}W^{(2)})-{1\over 2}(W^{(1)}DW^{(2)})
&=&-{5\over 3}\sqrt\hbar(W^{(1)}\partial W^{(3/2)})
-{1\over 3}\sqrt\hbar(DW^{(1)}DW^{(3/2)})\nonumber\\
&&+{1\over 4}\hbar^{3/2}\partial^2 DW^{(1)}\nonumber
\end{eqnarray}
In addition, there are several more dependent identities which we do not give
here.

We see that for $\hbar\ne 0$ and $\lambda=1/2$ the
identities (\ref{eq:qindep1/2}) can be used to solve for
$W^{(3)}$ and $W^{(7/2)}$ in terms of product of lower-spin
currents. On the other hand, for $\hbar\ne 0$ and $\lambda=0$
the identities (\ref{eq:qindep0}) can be applied to solve
for $W^{(s)}$ with $2\le s \le 7/2$. We therefore conclude that,
both for $\lambda=0$ as well as $\lambda=1/2$ there exists a
quantum telescoping procedure, which does not introduce an $s=1/2$ generator.

In the case $\lambda=1/2$ the construction leads to a
$c=-3$ representation of an $N=2$ super-$W_3$ algebra with as
independent currents the set $\{W^{(1)},W^{(3/2)},W^{(2)},W^{(5/2)}\}$.
{}From the OPE expansions given in the Appendix we see that the
operator products of these currents only give rise to $W^{(3)}$ and
$W^{(7/2)}$ as new operators. We use the identities (\ref{eq:qindep1/2})
to solve for these operators as composite
expressions in terms of the independent generators of the
$N=2$ super-$W_3$ algebra.
The resulting OPE expansions for the currents can be easily derived by
substituting $\lambda=1/2$ in the OPE expansions given in the Appendix.
In the resulting expressions it is understood that
everywhere $W^{(3)}$ and $W^{(7/2)}$
are given by the nonlinear expressions (\ref{eq:qindep1/2}).
We should note that the
$\{W^{(1)},W^{(3/2)}\}$ currents form a $c=-3$
representation of a $N=2$ super-Virasoro algebra.

A similar construction can be performed for the $\lambda=0$ case.
Like in the $\lambda=1/2$ case, we
take as independent currents the set $\{W^{(s)}\}$ with $1\le s\le 5/2$.
We now use the OPE expansions given in the Appendix for $\lambda=0$.
Again, these OPE's only give rise to $W^{(3)}$ and $W^{(7/2)}$
as independent currents. Next, we use the third and fourth identity
in (\ref{eq:qindep0}) to solve for these operators.
The $\lambda=0$ construction thus leads to a $c=3$ representation
of another $N=2$ super-$W_3$
algebra. To distinguish it from the $\lambda=1/2$ case we will call this
algebra $N=2$ super-$W_3'$.

A few remarks are in order.
First of all, we
should note that there is an arbitrariness in the way we
decide to write the $W^{(3)}$ and $W^{(7/2)}$ currents as composite
operators. This is due to the fact that there are more representation-dependent
relations at spin-3 and spin-7/2, than the ones which are used
to solve for the $W^{(3)}$ and $W^{(7/2)}$ operators (see
eqs.~(\ref{eq:nonl1}), (\ref{eq:nonl2}) for $\lambda=1/2$ and
eq.~(\ref{eq:qindep})
for $\lambda=0$).
One may always add to a given decomposition these
identities multiplied by an arbitrary coefficient. The reason
that this arbitrariness occurs is that we are working in the
context of a special two-scalar superfield realisation at
$c=-3, c=+3$ for $\lambda=1/2, \lambda=0$, respectively.
Therefore, our result may only be viewed as the $c=-3\ (c=+3)$ realisation
of a $N=2$ super-$W_3$\
($N=2$ super-$W_3'$) algebra at arbitrary $c$ modulo the special
$c=-3\ (c=+3)$ identities mentioned above.
To fix the arbitrariness one should use a representation
based upon $2i\ (i\ge 2)$ scalar superfields:

\begin{equation}
\{\phi_i,\bar\phi_i\}\hskip 2cm i\ge 2
\end{equation}
These fields should provide a representation of the algebra at
$\underline {\rm arbitrary}$ value of the central charge.
For instance, the $i=2$ representation
of the $N=2$ super-$W_3$ algebra was given recently in \cite{Ne,It}.
More recently, using a technique of \cite{Ro2}, a representation
of the $N=2$ super-$W_3$ algebra for
arbitrary values of $i$ has been given \cite{Lu4}.

Secondly,
in the $i=1$ representation
the supercurrents $W^{(2)}$ and $W^{(5/2)}$ are not primary
with respect to the $N=1$ super-Virasoro algebra generated
by $W^{(3/2)}$ (they are quasi-primary though).
The question now is whether we can make these
currents primary by an appropriate (nonlinear)
redefinition. In eqs.~(\ref{eq:red}), (\ref{eq:red1/2}) it is shown
that such a redefinition
is possible for arbitrary value of $\lambda$.
For $\lambda=1/2$ the supercurrents $\{W^{(2)},
W^{(5/2)}\}$ can be made $N=2$ primary with respect to the $N=2$
super-Virasoro algebra generated by $\{W^{(1)}, W^{(3/2)}\}$.
More explicitly,
for $\lambda=1/2$ the redefinitions (\ref{eq:red}), (\ref{eq:red1/2})
can be written in
terms of the generators of the $N=2$ super-$W_3$ algebra\
(i.e.~without $W^{(1/2)}$) as follows:

\begin{eqnarray}
\label{eq:qpp}
W^{(2)'}&=& 4 W^{(2)}-(W^{(1)}W^{(1)})+{2\over 3}D W^{(3/2)}\\
W^{(5/2)'} &=& 3W^{(5/2)} - 2 (W^{(1)}W^{(3/2)})\nonumber
\end{eqnarray}
It turns out that
for $\lambda=0$ the redefinitions (\ref{eq:red}) involve
the
$s=1/2$ generator.
Therefore, for $\lambda=0$, the algebra contains
quasi-primary generators
that cannot be made primary by a suitable redefinition.
This shows that the $\lambda=0$ and $\lambda=1/2$ cases lead two
inequivalent algebras.

Finally, we note that in general it might happen that the particular value of
the central charge we are working with corresponds to a singularity in
the expressions for the super-$W$ algebra at arbitrary $c$\footnote
{We thank C.~Pope and K.~Schoutens for a discussion on the issue
raised in this paragraph.}. For instance,
in the $N=2$ super-$W_3$ algebra of \cite{Ro1} a factor of
${1\over c+3}$ occurs. Therefore, for $c=-3$ some of the
expressions become singular\footnote{
The singularity at $c=-3$ is also discussed in \cite{Ro1}.}.
Since the results of \cite{Ro1}
are given in a primary basis it is difficult to compare.
We expect that in the $c=-3$
representation we are using all the expressions multiplying a
${1\over c+3}$ factor
are zero identically. Indeed we note that there are
nontrivial nonlinear identities at spin 3 and 7/2 (see eqs.~(\ref{eq:nonl1})
and (\ref{eq:nonl2})).
It would be interesting to verify whether this is indeed the case.

\vspace{.5cm}

\noindent{\bf 8. Conclusions}

\vspace{.5cm}

In this paper we have quantised the classical gauge theory of
$N=2\ w_\infty$-supergravity. We have used a representation in terms of
two scalar superfields corresponding to a two-dimensional target space
with Minkowskian signature. Like in the bosonic case we find that
the underlying classical algebra deforms into a corresponding
quantum algebra, thereby removing all matter-dependent anomalies.
The universal anomalies can be cancelled as well by choosing an appropriate
matter system.

We have shown how our results
can be used to obtain a two-scalar superfield realisation
of quantum $W_3$-supergravity. We gave the explicit answer for the case of
quantum $N=2\ W_3$-supergravity at $c=-3$ and
quantum $N=2$ super-$W_3'$-supergravity at $c=+3$.
In the first case all generators of the underlying algebra
can be made primary and,
although we have not explicitly checked this, we expect that
this case provides a $c=-3$ realisation of the $N=2$
super-$W_3$ algebra given in \cite{Ro1}.
In the latter case the underlying algebra contains quasi-primary
generators that cannot be made primary by some suitable redefinition.
It would be interesting to
investigate whether or not this case corresponds to the $c=+3$ realisation
of a $N=2$ super-$W_3'$ algebra at arbitrary values of $c$.
In this context it is
interesting to note that the occurrence of an $N=2$
super-$W$ algebra with quasi-primary generators
is also mentioned in the work of \cite{Ko}.

It would be interesting to extend out work to the general case
of quantum $N=2\ W_N$-supergravity theories. We expect that
the $N=3$ decomposition rules described in this paper should also work
for $N > 3$.
One question that arises here is whether the higher-spin generators can be
made $N=2$ primary for the same value of the parameter
$\lambda$.

An unusual feature we encountered is that
the telescoping procedure needed to perform the truncation to
super-$W_N$ only exists at the quantum level but not at the classical level.
We note that the situation is different in the bosonic case.
The difference can be understood from the following dimension
counting argument. Consider the schematic form of the currents
in the bosonic and supersymmetric case:

\begin{eqnarray}
w^{(s)} &\sim& (\partial\phi)^s\hskip 3cm {\rm bosonic}\\
w^{(s)} &\sim& (\partial\phi)^{s-1}D\phi D\bar\phi\hskip
1.5cm {\rm supersymmetric}\nonumber
\end{eqnarray}
{}From these expressions we see that in the bosonic case $w^{(s)}$ and
$w^{(t)}w^{(u)}$ with $s=t+u$ have the same dimensions (we remind that
the dimension of $\phi$ is $\sqrt\hbar$). However, in the
supersymmetric case $w^{(s)}$ has dimension $(s+1)\sqrt\hbar$ whereas
$w^{(t)}w^{(u)}$ has dimension $(s+2)\sqrt\hbar$. Therefore, in the
supersymmetric case one can only have identities of the form
$w^{(s)}=1/\sqrt\hbar w^{(t)}w^{(u)}$ which only makes sense
at the quantum level.
Due to the $1/\sqrt\hbar$ factors,
using the two-scalar superfield realisation of
quantum $W_N$-supergravity, one cannot take the classical limit
to obtain a realisation of a classical $w_N$-supergravity theory.
Only for $N=\infty$ a classical supergravity theory in terms of
commuting superfields (allowing a superstring coordinate interpretation)
can be defined.
This raises the issue of how to define the classical limit
of a quantum $W_N$ superstring theory.

\vfill\eject

\centerline{\bf Acknowledgements}

\vspace{.3cm}

We would like to thank M.~Blau,
J.~de Boer, C.~Pope and K.~Schoutens for discussions.
For one of us (E.B.) this work has been made possible by a
fellowship of the Royal Netherlands Academy of Arts and Sciences (KNAW).

\vfill\eject
\noindent{\bf Appendix A: Currents and OPE's}

\vspace{.5cm}

In this appendix we give the explicit forms of all
currents $W_\lambda^{(s)}$ with $1/2 \le s \le 5/2$ and
the (singular part of the) operator product expansions between them. These
expressions can be derived by a straightforward application of the
general formulae given in section 5.

The first few supercurrents take the form

\begin{eqnarray}
\label{eq:example}
W_\lambda^{(1/2)} &= & BC\nonumber \\
W_\lambda^{(1)}& = & (1-2\lambda)(DB)C - 2\lambda B(DC) \nonumber \\
W_\lambda^{(3/2)}& = & {1\over 2}(1-2\lambda)(\partial B)C
- {1\over 2}(DB)(DC) - \lambda
B(\partial C)  \nonumber \\
W_\lambda^{(2)} &= &{1\over 3}(1-2\lambda)(1-\lambda) (\partial DB)C
-{1\over 3}(1+2\lambda)(1-\lambda)(\partial B)(DC) \nonumber \\
&&-{1\over 3}(1+2\lambda)(1-\lambda)(DB)(\partial C)
+{1\over 3}\lambda(2\lambda+1) B(\partial DC) \\
W_\lambda^{(5/2)}& = & {1\over 6}(1-2\lambda)(1-\lambda)(\partial^2B) C
-{1\over 3}(1-\lambda)(\partial DB)(DC) \nonumber \\
&&-{1\over 3}(1+2\lambda)(1-\lambda)(\partial B)(\partial C)
+{1\over 6}(1+2\lambda )(DB)(\partial DC)
+ {1\over 6}\lambda(1+2\lambda) B(\partial^2 C) \nonumber \\
&&\vdots   \nonumber
\end{eqnarray}
The operator product expansions between these currents
are given by (we set $\hbar=1$)

\begin{eqnarray}
W_\lambda^{(1/2)}(1)W_\lambda^{(1/2)}(2) &\sim&
\ \ \ {\rm regular}
\nonumber\\
W_\lambda^{(1/2)}(1)W_\lambda^{(1)}(2) &\sim&
{W_\lambda^{(1/2)}\over z_{12}} + {\theta_{12}\over z_{12}^2}
\ \ \ +\ \ \ {\rm regular}
\nonumber\\
W_\lambda^{(3/2)}(1)W_\lambda^{(1/2)}(2) &\sim&
\{{1\over 2}{\theta_{12}W_\lambda^{(1/2)}\over z_{12}^2}
-{1\over 2} {D_2W_\lambda^{(1/2)}\over z_{12}}
+ {\theta_{12}\partial_2W_\lambda^{(1/2)}\over z_{12}}\}
-{1\over 2}{1\over z_{12}^2}
\ \ \ +\ \ \ {\rm regular}
\nonumber\\
W_\lambda^{(1/2)}(1)W_\lambda^{(2)}(2) &\sim&
{W_\lambda^{(3/2)}\over z_{12}} + \{{\theta_{12}W_\lambda^{(1)}
\over z_{12}^2} - {1\over 2}{D_2W_\lambda^{(1)}\over z_{12}}\}
\nonumber\\
&&-{4\over 3}(\lambda -{1\over 4})\{ {W_\lambda^{(1/2)}\over z_{12}^2}
-{\theta_{12}D_2W_\lambda^{(1/2)}\over z_{12}^2}
 - {\partial_2 W_\lambda^{(1/2)}
\over z_{12}} \}
\nonumber\\
&&- {4\over 3}(\lambda-{1\over 4}) {\theta_{12}\over z_{12}^3}
\ \ \ +\ \ \ {\rm regular}
\nonumber\\
W_\lambda^{(1/2)}(1)W_\lambda^{(5/2)}(2) &\sim&
{1\over 2}\{3{\theta_{12}W_\lambda^{(3/2)}\over z_{12}^2}
- {D_2W_\lambda^{(3/2)}\over z_{12}}\}
\nonumber\\
&&+{1\over 4}\{2{W_\lambda^{(1)}\over z_{12}^2}
-{\theta_{12}D_2W_\lambda^{(1)}\over z_{12}^2}
-{\partial_2W_\lambda^{(1)}\over z_{12}}\}
\nonumber\\
&&-{1\over 3}(\lambda - {1\over 4})\{4{\theta_{12}W_\lambda^{(1/2)}
\over z_{12}^3}
-4{D_2W_\lambda^{(1/2)}\over z_{12}^2}
-4{\theta_{12}\partial_2W_\lambda^{(1/2)}\over z_{12}^2}
+2{\partial_2D_2W_\lambda^{(1/2)}\over z_{12}^3}\}
\nonumber\\
&&-{2\over 3}(\lambda - {1\over 4}) {1\over z_{12}^3}
\ \ \ +\ \ \ {\rm regular}
\nonumber\\
W_\lambda^{(1)}(1)W_\lambda^{(1)}(2) &\sim&
-2{\theta_{12}W_\lambda^{(3/2)}\over z_{12}} -4 {\lambda - {1\over 4}
\over z_{12}^2}
\ \ \ + \ \ \ {\rm regular}
\\
W_\lambda^{(3/2)}(1)W_\lambda^{(1)}(2) &\sim&
\{{\theta_{12}W_\lambda^{(1)}\over z_{12}^2} - {1\over 2}{D_2W_\lambda^{(1)}
\over z_{12}} + {\theta_{12}\partial_2W_\lambda^{(1)}\over z_{12}}\}
\ \ \ +\ \ \ {\rm regular}
\nonumber\\
W_\lambda^{(1)}(1)W_\lambda^{(2)}(2) &\sim&
-2 {\theta_{12}W_\lambda^{(5/2)}\over z_{12}}
-{8\over 3}(\lambda-{1\over 4}){W_\lambda^{(1)}\over z_{12}^2}
\nonumber\\
&&+4\lambda(\lambda-{1\over 2})\{
{\theta_{12}W_\lambda^{(1/2)}\over z_{12}^3}
-{D_2W_\lambda^{(1/2)}\over z_{12}^2}\}
+4{\lambda(\lambda-{1\over 2})\over z_{12}^3}
\ \ \ +\ \ \ {\rm regular}
\nonumber\\
W_\lambda^{(1)}(1)W_\lambda^{(5/2)}(2) &\sim&
-{1\over 2}\{
4{\theta_{12}W_\lambda^{(2)}\over z_{12}^2}
-{D_2W_\lambda^{(2)}\over z_{12}}
+{\theta_{12}\partial_2W_\lambda^{(2)}\over z_{12}}\}
\nonumber\\
&&-{8\over 3}(\lambda -{1\over 4}){W_\lambda^{(3/2)}\over z_{12}^2}
\nonumber\\
&&+2\lambda(\lambda -{1\over 2})\{{W_\lambda^{(1/2)}\over z_{12}^3}
-{\theta_{12}D_2W_\lambda^{(1/2)}\over z_{12}^3}
-{\partial_2W_\lambda^{(1/2)}\over z_{12}^2}\}
\nonumber\\
&&+2\lambda(\lambda-{1\over 2}){\theta_{12}\over z_{12}^4}
\ \ \ +\ \ \ {\rm regular}
\nonumber\\
W_\lambda^{(3/2)}(1)W_\lambda^{(3/2)}(2) &\sim&
\{{3\over 2}{\theta_{12}W_\lambda^{(3/2)}\over z_{12}^2}
-{1\over 2}{D_2W_\lambda^{(3/2)}\over z_{12}}
+{\theta_{12}\partial_2 W_\lambda^{(3/2)}\over z_{12}}\}
+ 2{\lambda - {1\over 4}\over z_{12}^3}
\ \ \ + \ \ {\rm regular}
\nonumber\\
W_\lambda^{(3/2)}(1)W_\lambda^{(2)}(2) &\sim&
\{2{\theta_{12}W_\lambda^{(2)}\over z_{12}^2}
-{1\over 2}{D_2W_\lambda^{(2)}\over z_{12}}
+{\theta_{12}\partial_2W_\lambda^{(2)}\over z_{12}}\}
\nonumber\\
&&-2\lambda(\lambda-{1\over 2}){W_\lambda^{(1/2)}\over z_{12}^3}
-2\lambda(\lambda-{1\over 2}){\theta_{12}\over z_{12}^4}
\ \ \ + \ \ \ {\rm regular}
\nonumber\\
W_\lambda^{(3/2)}(1)W_\lambda^{(5/2)}(2) &\sim&
\{
{5\over 2}{\theta_{12}W_\lambda^{(5/2)}\over z_{12}^2}
-{1\over 2}{D_2W_\lambda^{(5/2)}\over z_{12}}
+{\theta_{12}\partial_2W_\lambda^{(5/2)}\over z_{12}}\}
\nonumber\\
&&+{4\over 3}(\lambda - {1\over 4}){W_\lambda^{(1)}\over z_{12}^3}
\nonumber\\
&&-3\lambda(\lambda -{1\over 2})\{{\theta_{12}W_\lambda^{(1/2)}\over z_{12}^4}
-{D_2W_\lambda^{(1/2)}\over z_{12}^3}\}
-3\lambda(\lambda -{1\over 2}){1\over z_{12}^4}
\ \ \ +\ \ \ {\rm regular}
\nonumber\\
W_\lambda^{(2)}(1)W_\lambda^{(2)}(2) &\sim&
-2{\theta_{12}W_\lambda^{(7/2)}\over z_{12}}
\nonumber\\
&&-{8\over 3}(\lambda -{1\over 4})\{
{W_\lambda^{(2)}\over z_{12}^2}
+{1\over 2}{\theta_{12}D_2W_\lambda^{(2)}\over z_{12}^2}
+{1\over 2}{\partial_2W_\lambda^{(2)}\over z_{12}}
+{3\over 10}{\theta_{12}\partial_2D_2W_\lambda^{(2)}\over z_{12}}\}
\nonumber  \\
&&+{4\over 3}(\lambda-1)(2\lambda+1)\{
{\theta_{12}W_\lambda^{(3/2)}\over z_{12}^3}
-{1\over 3}{D_2 W_\lambda^{(3/2)}\over z_{12}^2}
+{2\over 3}{\theta_{12}\partial_2W_\lambda^{(3/2)}\over z_{12}^2}
\nonumber\\
&&-{1\over 6}{\partial_2D_2W_\lambda^{(3/2)}\over z_{12}}
+{1\over 4}{\theta_{12}\partial_2^2W_\lambda^{(3/2)}\over z_{12}}\}
\nonumber\\
&&-{4\over 3}(\lambda-{1\over 4}){1+2\lambda-4\lambda^2\over z_{12}^4}
\ \ \ + \ \ \  {\rm regular}
\nonumber\\
W_\lambda^{(2)}(1)W_\lambda^{(5/2)}(2) &\sim&
\{-3{\theta_{12}W_\lambda^{(3)}\over z_{12}^2}
+{1\over 2}{D_2W_\lambda^{(3)}\over z_{12}}
-{3\over 2}{\theta_{12}\partial_2W_\lambda^{(3)}\over z_{12}}\}
\nonumber\\
&&-{4\over 3}(\lambda - {1\over 4})\{
{W_\lambda^{(5/2)}\over z_{12}^2}
+{2\over 5}{\theta_{12}D_2W_\lambda^{(5/2)}\over z_{12}^2}
+{2\over 5}{\partial_2W_\lambda^{(5/2)}\over z_{12}}
+{1\over 5}{\theta_{12}\partial_2D_2W_\lambda^{(5/2)}\over z_{12}}\}
\nonumber\\
&&+{1\over 6}(\lambda -1)(2\lambda +1)\{
4{\theta_{12}W_\lambda^{(1)}\over z_{12}^4}
-2{D_2W_\lambda^{(1)}\over z_{12}^3}
+2{\theta_{12}\partial_2W_\lambda^{(1)}\over z_{12}^3}
\nonumber\\
&&-{2\over 3}{\partial_2D_2W_\lambda^{(1)}\over z_{12}^2}
+{2\over 3}{\theta_{12}\partial_2^2W_\lambda^{(1)}\over z_{12}^2}
-{1\over 6}{\partial_2^2D_2W_\lambda^{(1)}\over z_{12}}
+{1\over 6}{\theta_{12}\partial_2^3W_\lambda^{(1)}\over z_{12}}\}
\nonumber\\
&&+\ \ \ {\rm regular}\nonumber\\
W_\lambda^{(5/2)}(1)W_\lambda^{(5/2)}(2) &\sim&
\{{7\over 2}{\theta_{12}W_\lambda^{(7/2)}\over z_{12}^2}
-{1\over 2}{D_2W_\lambda^{(7/2)}\over z_{12}}
+2{\theta_{12}\partial_2W_\lambda^{(7/2)}\over z_{12}}\}
\nonumber\\
&&+(\lambda - {1\over 4})\{
{4\over 3}{W_\lambda^{(2)}\over z_{12}^3}
+{2\over 3}{\theta_{12}D_2W_\lambda^{(2)}\over z_{12}^3}
+{2\over 3}{\partial_2W_\lambda^{(2)}\over z_{12}^2}
\nonumber\\
&&+{2\over 5}{\theta_{12}\partial_2D_2W_\lambda^{(2)}\over z_{12}^2}
+{1\over 5}{\partial_2^2W_\lambda^{(2)}\over z_{12}}
+{2\over 15}{\theta_{12}\partial_2^2D_2W_\lambda^{(2)}\over z_{12}}\}
\nonumber\\
&&+{1\over 6}(\lambda -1)(2\lambda +1)\{
-10{\theta_{12}W_\lambda^{(3/2)}\over z_{12}^4}
+{10\over 3}{D_2W_\lambda^{(3/2)}\over z_{12}^3}
-{20\over 3}{\theta_{12}\partial_2W_\lambda^{(3/2)}\over z_{12}^3}
\nonumber\\
&&+{5\over 3}{\partial_2D_2W_\lambda^{(3/2)}\over z_{12}^2}
-{5\over 2}{\theta_{12}\partial_2^2W_\lambda^{(3/2)}\over z_{12}^2}
+{1\over 2}{\partial_2^2D_2W_\lambda^{(3/2)}\over z_{12}}
-{2\over 3}{\theta_{12}\partial_2^3W_\lambda^{(3/2)}\over z_{12}}\}
\nonumber\\
&&+{4\over 3}(\lambda-{1\over 4}){1+2\lambda-4\lambda^2\over z_{12}^5}
\ \ \ +\ \ \ {\rm regular}\nonumber
\end{eqnarray}
We have only given one order of the currents in each OPE. The
expression for the other order can be easily derived by using the
identity
$W_\lambda^{(s)}(1)W_\lambda^{(t)}(2) = (-)^{|2s|_2|2t|_2}
W_\lambda^{(t)}(2)W_\lambda^{(s)}(1)$ and the super-Taylor expansion rule.

\vfill\eject


\begin{thebibliography}{99}
\bibitem{Za} A.B.~Zamolodchikov, Teor.~Mat.~Fiz.~{\bf 65} (1985) 347.
\bibitem{Fa} V.A.~Fateev and A.B.~Zamolodchikov, Nucl.~Phys.~{\bf B280\
[FS18]} (1987) 644;
V.A.~Fateev and S.~Lukyanov, Int.~J.~Mod.~Phys.~{\bf A3}
(1988) 507.
\bibitem{Ba} F.~Bais, P.~Bouwknegt, M.~Surridge and K.~Schoutens,
Nucl.~Phys.~{\bf B304} (1988) 348, 371.
\bibitem{Bi1} A.~Bilal and J.-L.~Gervais, Nucl.~Phys.~{\bf B314} (1989) 646;
{\bf B318} (1989) 579.
\bibitem{Bak} I.~Bakas, Commun.~Math.~Phys.~{\bf 134} (1990) 487.
\bibitem{Po1} C.N.~Pope, L.J.~Romans and X.~Shen, Phys.~Lett.~{\bf 236B}
(1990) 173.
\bibitem{Po2} C.N.~Pope, L.J.~Romans and X.~Shen, Nucl.~Phys.~{\bf B339} (1990)
191.
\bibitem{Se} E.~Sezgin and E.~Sokatchev, Phys.~Lett.~{\bf B227} (1989) 103;
E.~Sezgin, in {\it Strings '89} (World Scientific, 1990).
\bibitem{Po3} C.N.~Pope and X.~Shen, Phys.~Lett.~{\bf B236} (1990) 21;
in {\it High Energy Physics and Cosmology}, Proceedings of the
1989 Trieste Summer School (World Scientific, 1990).
\bibitem{Be3} E.~Bergshoeff, C.N.~Pope, L.J.~Romans, E.~Sezgin and
X.~Shen, Phys.~Lett.~{\bf 245B} (1990) 447.
\bibitem{In} T.~Inami, Y.~Matsuo and I.~Yamanaka, Phys.~Lett.~{\bf 215B}
(1988) 701; Int.~J.~Mod.~Phys.~{\bf A5} (1990) 4441.
\bibitem{Ho} K.~Hornfeck and E.~Ragoucy, Nucl.~Phys.~{\bf B340} (1990) 225;
A.~Bilal, Phys.~Lett.~{\bf 238B} (1990) 239;
K.~Schoutens and A.~Sevrin, Phys.~Lett.~{\bf 258B} (1991) 134.
\bibitem{Fi}J.~Figueora-O'Farrill and
S.~Schrans, "The conformal bootstrap and super $W$-algebras'', preprint
KUL-TF-90/16, Int.~Journ.~Mod.~Phys.~A, to be published;
Phys.~Lett.~{\bf 257B} (1991) 69.
\bibitem{No} K.~Mohri and H.~Nohara, Nucl.~Phys.~{\bf B349} (1991)
253.
\bibitem{Ko}
S.~Komata, K.~Mohri and H.~Nohara, Nucl.~Phys.~{\bf B359} (1991) 168.
\bibitem{Ah}C.~Ahn,
K.~Schoutens and A.~Sevrin, Int.~J.~Mod.~Phys.~{\bf A6} (1991) 3467.
\bibitem{Lu2} H.~Lu, C.N.~Pope, L.J.~Romans, X.~Shen and X-J.~Wang,
Phys.~Lett.~{\bf B264} (1991) 91.
\bibitem{Ro1} L.J.~Romans, ``The $N=2$ super-$W_3$ algebra'',
preprint, USC-91/HEP06.
\bibitem{Ne}D.~Nemeschansky and S.~Yankielowicz, ``$N=2\ W$-algebras,
Kazama-Suzuki models and Drinfeld-Sokolov reduction'', preprint,
USC-91/005.
\bibitem{It} K.~Itoh, ``$N=2$ Superconformal $CP_n$ Model'',
preprint, YITP/K-934 (June 1991).
\bibitem{Fu1}
M.~Fukuma, H.~Kawai and R.~Nakayama, Int~J.~Mod.~Phys.~{\bf A6}
(1991) 1385;
R.~Dijkgraaf, H.~Verlinde and E.~Verlinde, Nucl.~Phys.~{\bf B348}
(1991) 435.
\bibitem{Ya} K.~Yamagishi, Phys.~Lett.~{\bf 259B} (1991) 436;
F.~Yu and Y.-S.~Wu, Phys.~Lett.~{\bf 263B} (1991) 220.
\bibitem{Go} J.~Goeree, Nucl.~Phys.~{\bf B358} (1991) 737.
\bibitem{Fu2}M.~Fukuma, H.~Kawai and R.~Nakayama,  "Infinite-dimensional
Grassmannian structure of two-dimensional quantum gravity'', preprint,
UT-572; KEK-TH-272 (November 1990).
\bibitem{Kl} I.R.~Klebanov and A.~M.~Polyakov,
``Interaction of Discrete States in Two-dimensional String Theory'',
preprint, PUPT-1281 (September 1991).
\bibitem{El} J.~Ellis, N.E.~Mavromatos and D.V.~Nanopoulos,
``On the connection between Quantum Mechanics and the geometry
of two-dimensional strings'', preprint, ACT-45, CERN-TH.6229/91,
CTP-TAMU-66/91 (September 91).
\bibitem{Av} J.~Avan and A.~Jevicki, ``String Field Actions
from $W_\infty$'', preprint, BROWN-HET-839 (October 1991).
\bibitem{Hu1} C.M.~Hull, Phys.~Lett.~{\bf 240B} (1989) 110;
Nucl.~Phys.~{\bf B353} (1991) 107.
\bibitem{Sc}
K.~Schoutens, A.~Sevrin and P.~van Nieuwenhuizen,
Phys.~Lett.~{\bf 243B} (1990) 245;
Phys.~Lett.~{\bf 251B} (1990) 355; Nucl.~Phys.~{\bf B349} (1991) 791.
\bibitem{Be1}E.~Bergshoeff, C.N.~Pope, L.J.~Romans, E.~Sezgin, X.~Shen
and K.S.~Stelle, Phys.~Lett.~{\bf 243B} (1990) 350.
\bibitem{Be6} E.~Bergshoeff,
C.N.~Pope and K.S.~Stelle, Phys.~Lett.~{\bf 249B} (1990) 208.
\bibitem{Ma}
Y.~Matsuo, Phys.~Lett.~{\bf 227B} (1989) 222.
\bibitem{Li}
K.~Li and C.N.~Pope, Class.~Quantum Grav.~{\bf 8} (1991) 1677.
\bibitem{Hu5}
C.M.~Hull, ``W-gravity anomalies 1; Induced quantum W gravity'',
preprint, QMW/PH/91/2;
W-gravity anomalies 2; Matter-dependent anomalies of non-linearly
realised symmetries, preprint, QMW/PH/91/3.
\bibitem{Sc2}
K.~Schoutens, A.~Sevrin and P.~van Nieuwenhuizen, Nucl.~Phys.~{\bf B364}
(1991) 584;
Loop-calculations in BRST-quantised chiral $W_3$ gravity'',
preprint, ITP-SB-91-13.
\bibitem{Be2} E.~Bergshoeff, P.S.~Howe, C.N.~Pope, E.~Sezgin, X.~Shen
and K.S.~Stelle, Nucl.~Phys.~{\bf B363} (1991) 163.
\bibitem{Ce}
A.T.~Ceresole, M.~Frau, J.~McCarthy and A.~Lerda,
Phys.~Lett.~{\bf 265B} (1991) 72.
\bibitem{Po5}
C.N.~Pope, L.J.~Romans and K.S.~Stelle, Phys.~Lett.~{\bf 268B} (1991) 167.
\bibitem{Po4} C.N.~Pope, ``Anomaly-free $W$-gravity Theories'',
to appear in the proceedings of {\it Strings and Symmetries}, Stony Brook,
1991, preprint, CTP TAMU-82/91 (October 1991).
\bibitem{Da}
S.~Das, A.~Dhar and S.~Rama,
``Physical Properties of $W$ Gravities and $W$ Strings'', preprint,
TIFR/TH/91-11 (Februari 1991);
``Physical States and
Scaling Properties of $W$ gravities and $W$ Strings'', preprint,
Tata Institute 90-21 (March 1991).
\bibitem{Po7}
C.N.~Pope, L.J.~Romans and K.S.~Stelle,
Phys.~Lett.~{\bf 269B} (1991) 287;
C.N.~Pope, L.J.~Romans, E.~Sezgin and K.S.~Stelle, ``The
$W_3$ spectrum'', preprint, CTP TAMU-68/91, Imperial/TP/90-91/40,
IC/91/241 (September 1991).
\bibitem{Fig} J.~Figueora-O'Farrill, J.~Mas and E.~Ramos, ``Integrability
and BiHamiltonian structure of the even order SKdV hierarchies'',
preprint, KUL-TF-91/17 (April 1991).
\bibitem{Yu} F.~Yu, ``The Super KP Origin of Super $W_{1+\infty}$ Algebra
and Its Topological Version'', preprint, UU-HEP-91/12 (August 1991).
\bibitem{Das} A.~Das, E.~Sezgin and S.J.~Sin, ``The Super $W_\infty$
Symmetry of the Manin-Radul Super KP Hierarchy'', preprint,
UR-1232,ER-13065-685, CTP-TAMU-54/91, IC/91/383, UFIFT-HEP-91-26
(November 1991).
\bibitem{Lu3} H.~Lu, C.N.~Pope and X.~Shen, Nucl.~Phys.~{\bf B366}
(1991) 95.
\bibitem{Li2} K.~Li, ``Linear $w_N$ Gravity'', preprint, CALT-68-1724
(April 1991).
\bibitem{Bas} F.~Bastianelli, Mod.~Phys.~Lett.~{\bf A6} (1991) 425.
\bibitem{Mi}
A.~Mikovi\' c, Phys.~Lett.~{\bf B260} (1991) 75.
\bibitem{Be5} E.~Bergshoeff, C.N.~Pope, L.J.~Romans, E.~Sezgin and
X.~Shen, Mod.~Phys.~Lett.~A, Vol.~5, No.~24 (1990) 1957.
\bibitem{Be8} E.~Bergshoeff and M.~de Roo, ``$N=2\ w_\infty$-supergravity'',
preprint, UG-62/91 (September 1991).
\bibitem{Sh} X.~Shen and X.J.~Wang, ``Bosonisation of the Complex-Boson
Realisations of $W_\infty$'', preprint, CTP TAMU-59/91 (August 1991).
\bibitem{Hu6} C.M.~Hull, Nucl.~Phys.~{\bf B364} (1991) 621.
\bibitem{Be4} E.~Bergshoeff, B.~de Wit and M.~Vasiliev,
Phys.~Lett.~{\bf 256B} (1991) 199; "The structure of the super-$W_\infty
(\lambda)$ algebra, preprint, CERN.TH-6021-91, THU-91/05,
Nucl.~Phys.~{\bf B}, in press.
\bibitem{Fr}D.~Friedan, E.~Martinec and S.~Shenker, Nucl.~Phys.~{\bf
B271} (1986) 93.
\bibitem{Bakas} I.~Bakas, B.~Khesin and E.~Kiritsis,
``The Logarithm of the Derivative Operator and Higher Spin Algebras
of $W_\infty$ type'', preprint, UCB-PTH-91/48, LBL-31303,
UMD-PP-92-36, (September 1991).
\bibitem{Bie} L.C.~Biedenharn and J.D.~Louck,
Ann.~Phys.~(NY) {\bf 63} (1971) 459.
\bibitem{Fra} E.S.~Fradkin and M.A.~Vasiliev, Ann.~Phys.~(NY)
{\bf 177} (1987) 63;
M.A.~Vasiliev, Fortsch.~Phys.~{\bf 36} (1988) 33.
\bibitem{Bou} P.~Bouwknegt, A.~Ceresole, P.~van Nieuwenhuizen and
J.~McCarthy, Phys.~Rev.~{\bf D40} (1989) 415.
\bibitem{Mar} E.~Martinec and G.~Sotkov, Phys.~Lett.~{\bf B208} (1988)
249.
\bibitem{Ger} A.~Gerasimov, A.~Levin and A.~Marshakov, Nucl.~Phys.~{\bf
B360} (1991) 537.
\bibitem{Ya2} K.~Yamagishi, Phys.~Lett.~{\bf 266B} (1991) 829.
\bibitem{Po6} C.N.~Pope, L.J.~Romans and X.~Shen, Phys.~Lett.~{\bf 254B}
(1991) 401.
\bibitem{Lu1} H.~Lu, C.N.~Pope, X.~Shen and X.J.~Wang, Phys.~Lett.~{\bf 267B}
(1991) 356.
\bibitem{Be7} E.~Bergshoeff, ``Nonlinear super-$W$ algebras at fixed
central charge'', preprint, UG-50/91 (July 1991), Phys.~Lett., in press.
\bibitem{Ro2} L.J.~Romans, Nucl.~Phys.~{\bf B352} (1991) 829.
\bibitem{Lu4} H.~Lu, C.N.~Pope, X.J.~Wang and K.W.~Xu,
``Anomaly freedom and Realisations for Super-$W_3$ Strings'',
preprint, CTP TAMU-85/91 (October 1991).



\end{thebibliography}
\end{document}